\documentclass[%
preprint,
 amsmath,amssymb,
 aps,
]{revtex4-2}

\usepackage{color}
\usepackage{graphicx}
\usepackage{dcolumn}
\usepackage{bm}
\usepackage{booktabs} 

\newcommand{\thYoung}{\theta_\textnormal{eq}}
\newcommand{\thInitial}{\theta_\textnormal{in}}
\newcommand{\LJWdepth}{\varepsilon_{\textnormal{wf}}}
\newcommand{\LJdepth}{\varepsilon_{\textnormal{LJ}}}
\newcommand{\LJWdepthWetting}{\varepsilon_{\textnormal{wf}}^{\textnormal{crit}}}
\newcommand{\FE}{\mathcal{F}}
\newcommand{\GrandPotential}{\Omega}

\newcommand{\Wa}{\textnormal{Wa}^0}

\newcommand{\nDensity}{n}
\newcommand{\nDensityV}{\nDensity_{\textnormal{vap}}}
\newcommand{\nDensityL}{\nDensity_{\textnormal{liq}}}
\newcommand{\nDensityCrit}{\nDensity_{\textnormal{crit}}}
\newcommand{\nDensityLV}{\nDensity_{\textnormal{lv}}}
\newcommand{\nDensityWV}{\nDensity_{\textnormal{wv}}}
\newcommand{\nDensityWL}{\nDensity_{\textnormal{wl}}}

\newcommand{\nDensityInitial}{\nDensity_{\textnormal{in}}}

\newcommand{\Vext}{V_{\textnormal{ext}}}
\newcommand{\chemPot}{\mu}

\newcommand{\surfaceTension}{\gamma}
\newcommand{\surfaceTensionLV}{\surfaceTension_{\textnormal{lv}}}

\newcommand{\Grad}{{\boldsymbol \nabla}}
\newcommand{\Div}{{\boldsymbol \nabla} \cdot}
\newcommand{\klamm}[1]{\left( #1 \right)}
\newcommand{\klammCurl}[1]{\left\{ #1 \right\}}
\newcommand{\vel}{{\bf u}}
\newcommand{\diff}[2]{\frac{\partial #1}{\partial #2}}

\newcommand{\shearViscosity}{\eta}
\newcommand{\bulkViscosity}{\zeta}
\newcommand{\shearViscosityL}{\shearViscosity_{\textnormal{liq}}}
\newcommand{\shearViscosityV}{\shearViscosity_{\textnormal{vap}}}
\newcommand{\bulkViscosityL}{\bulkViscosity_{\textnormal{liq}}}
\newcommand{\bulkViscosityV}{\bulkViscosity_{\textnormal{vap}}}

\newcommand{\CLPos}{y_{1,0}}
\newcommand{\IdMatrix}{{\bf I}}

\newcommand{\CLVel}{U_{\textnormal{CL}}}

\newcommand{\particleMass}{m}

\newcommand{\viscousHeatProduction}{\dot \varepsilon_v}
\newcommand{\viscousHeatProductionRescaled}{\dot\varepsilon_{v}^\ast}
\newcommand{\LJdiam}{\sigma}

\newcommand{\frictionCoefficientMKT}{\mu_{\textnormal{f}}}

\newcommand{\pos}{{\bf r}}
\newcommand{\dI}{\textnormal{d}}
\newcommand{\defi}{{\mathrel{\mathop:}=}}

\newcommand{\FEhs}{\FE_{\textnormal{HS}}}
\newcommand{\FEattr}{\FE_{\textnormal{attr}}}
\newcommand{\BHattr}{\phi_{\textnormal{attr}}}

\newcommand{\IrrevStressTensor}{{\boldsymbol \tau}}

\usepackage{xr}
\usepackage{xcite}
\usepackage{xr-hyper}

\begin{document}


\title{Hydrodynamic density-functional theory for the moving contact-line problem reveals fluid
structure and emergence of a spatially distinct pattern.}


\author{Andreas Nold}
\affiliation{Theory of Neural Dynamics, Max Planck Institute for Brain Research, 60438 Frankfurt am Main, Germany}

\author{Benjamin D.~Goddard}
\affiliation{The School of Mathematics and Maxwell Institute for Mathematical Sciences, The University of Edinburgh, Edinburgh, EH9 3FD, UK}

\author{David N.~Sibley}
\affiliation{Department of Mathematical Sciences, Loughborough University, Loughborough, LE11 3TU, UK}

\author{Serafim Kalliadasis}
\email{s.kalliadasis@imperial.ac.uk}
\affiliation{Department of Chemical Engineering, Imperial College London, London, SW7 2AZ, UK}

\date{\today}

\begin{abstract}
Understanding the nanoscale effects controlling the dynamics of a contact
line -- defined as the line formed at the junction of two fluid phases and
a solid -- has been a longstanding problem in fluid mechanics pushing
experimental and numerical methods to their limits. A major challenge is
the multiscale nature of the problem, whereby nanoscale phenomena manifest
themselves at the macroscale. To probe the nanoscale, not easily accessible
to other methods, we propose a reductionist model that employs elements
from statistical mechanics, namely dynamic-density-functional theory
(DDFT), in a Navier-Stokes-like equation -- an approach we name
hydrodynamic DDFT. The model is applied to an isothermal
Lennard-Jones-fluid with no slip on a flat solid substrate. Our
computations reveal fluid stratification with an oscillatory density
structure close to the wall and the emergence of two distinct regions as
the temperature increases: a region of compression on the vapor side of the
liquid-vapour interface and an effective slip region of large shear on the
liquid side. The compressive region spreads along the fluid interface at a
lengthscale that increases faster than the width of the fluid interface
with temperature, while the width of the slip region is bound by the
oscillatory fluid density structure and is constrained to a few particle
diameters from the wall. Both compressive and shear effects may offset
contact line friction, while compression in particular has a
disproportionately high effect on the speed of advancing contact lines at
low temperatures.
\end{abstract}

\pacs{Valid PACS appear here}

\maketitle

\section{Introduction}

\begin{figure}[t]
\centering
\includegraphics[width=12cm]{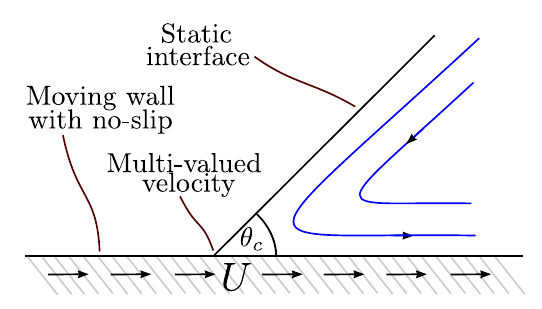}
\caption{Sketch of a stationary interface in contact with a wall moving with velocity $U$
with no slip. $\theta_c$ denotes the microscopic contact angle, i.e. the contact angle right
at the contact point. Classical hydrodynamic theory results in a multivaluedness of the velocity,
with the consequence that the hydrodynamic problem has no solution, and associated with this,
the shear stress and pressure are singular at the contact point with a $1/r$ singularity,
with $r$ the distance from the contact point (hence a non-integrable shear-stress/pressure
singularity)~\cite{Huh.19711}.\label{fig:Fig1_SketchMovingCA}}
\end{figure}

Wetting or dewetting -- the fluid mechanical process of a liquid phase
expanding or shrinking its contact area with a solid substrate -- is
ubiquitous in nature and technology \citep{natcomlh,*Koh_etal}. Central to
any description of wetting or dewetting is the presence of a moving contact
line, the line formed by the intersection of a fluid-fluid interface with a
solid substrate (see Fig. \ref{fig:Fig1_SketchMovingCA}). Classical fluid
mechanics predicts that the force at the contact line is infinite, an
unphysical result called the ``contact line
singularity"~\citep{Huh.19711,Bonn.20090527}. It is often concluded that
nanoscale effects, neglected in the standard continuous models, resolve the
singularity. However, as of yet it is unclear what the precise role of these
effects is and how they manifest to influence macroscale behavior.

Experimental, numerical and theoretical approaches have provided clues
towards possible solutions, but each approach faces its own unique
challenges. Experimental measurements cannot capture single particle dynamics
\cite{Rame2004,Seveno2009,Ralston:2008fk}. Numerical particle-based, i.e.
molecular dynamics (MD) simulations have allowed for valuable insights into
nanodroplet dynamics. They are, however, limited due to low signal-to-noise
ratios, also present in experimental techniques, prohibitive computational
costs \cite{Hadjiconstantinou}, theoretical assumptions imposed by
unconstrained approximations such as the thermostat choice as well as
boundary conditions~\cite{Antonio2019}, and high number of parameters and
degrees of freedom~\cite{smith2018moving}. For instance, a global thermostat,
such as Nos\'e-Hoover, for problems in the presence of shear can lead to
thermal energy production close to solid boundaries higher than the one in
the bulk, potentially causing a non-uniform temperature distribution in a
fluid, especially at low temperatures. On the theoretical front, one of the
earliest models, molecular kinetic theory (MKT) \cite{Blake.19697}, views
wetting as an activated chemical adsorption process. Its relation of contact
line friction to the off-equilibrium driving force is obtained as a
constitutive law following thermodynamic arguments \cite{ren2010continuum},
and was later used as a nanoscale boundary condition for the viscocapillary
regime~\cite{petrov1992combined,ren2010continuum}. [This region, where
viscous effects balance surface tension, is discussed in detail
in~\cite{Sibley:2015:CrackingAnOldNut} where the asymptotics of the
hydrodynamic moving contact line problem were carefully revisited
highlighting some misconceptions with certain technical details in the
asymptotic analysis of previous works, e.g. by Hocking and
Cox~\citep{Hocking:1983fk,*Cox.1986}.] Other approaches introduce time- and
density-dependent surface tensions \cite{Shikhmurzaev:1993uq}, or generalized
Navier boundary conditions which combine slip, viscous stress, and
uncompensated Young stress \cite{QianWangSheng:2003}. However, by design
these models include at least one phenomenological parameter and fall short
of identifying the precise nanoscale effects that determine the fluid
structure and compete to resolve the contact line singularity, and, not
surprisingly, drawing significant debate about model choices and validity
\cite{Sibley:2014:Comparison,EPST:Wetting,SnoeijerAndreotti:2013}.

To circumvent the experimental and numerical-theoretical limitations, we
follow a theoretical framework based on the statistical mechanics of
classical fluids, namely density-functional theory
(DFT)~\citep{Evans1979,*Lutsko2010}, which naturally includes particle-level
information. As highlighted by Lutsko~\cite{Lutsko2019}, the advantage of DFT
over other theories, e.g. diffuse interfaces, is that it is, in principle, ab
initio offering a quantitatively accurate representation of molecular-scale
correlations and structure; a fundamental foundational theory and from which
other theories can be viewed as approximations.
It has been successfully applied to highly non-uniform systems from wetting,
thin films and drops~\citep{PeterPRE,Peter2016,Buller2017,Peter2021}, crystal
structures-- crystallization~\citep{Lutsko2019,Schmidt2020} and
solidification~\cite{Archer2014}, to complex fluids such as
polymers~\citep{Wu2002,Roth2020} and lipid bilayers~\citep{Chapman2007}, and
molecular self-assembly~\citep{Wu2006}. DFT has been extended to dynamics,
the so-called dynamic DFT (DDFT), which describes not only equilibrium but
also the dynamic behavior out of equilibrium, and has been validated in a
wide variety of situations, showing excellent agreement with
stochastic/Langevin simulations
\citep{Archer:2009fk,Goddard:2013Unification,goddard2020wellposedness,teVrugt2020}.
Here DDFT will be applied to a problem that simulations cannot easily tackle.
In this direction we couple DDFT with a hydrodynamic model, a generalisation
which we refer to as hydrodynamic DDFT (HDDFT). It results in a reductionist
continuum model that depends on the fluid viscosity and three truly
fundamental parameters of the system: fluid temperature, fluid-fluid, and
wall-fluid interactions (these interactions in turn depend on the interatomic
potential). It retains information from the particle scale while also
capturing the viscocapillary region. It therefore enables scrutiny of the
molecular organisation of the moving contact line in the immediate vicinity
of the contact point. In particular, we can probe the nanoscale behavior as
the temperature varies, inaccessible with other experimental and numerical
approaches, such as MD, and so far remains largely unstudied.

To check consistency with meso- and macroscopic predictions and experimental
results, we show that HDDFT recovers a linear relationship between contact
line velocity and force, analogous to molecular kinetic theory (MKT), with
the contact line friction as a parameter \cite{Blake.19697}. HDDFT also
provides predictions for the contact line friction as a function of
temperature and wall attraction for hitherto unprobed values.

Zooming into the nanoscale region of the contact line, we observe l fluid
stratification with an oscillatory density structure close to the wall and
the emergence of two distinct regions: a shear region which extends along the
wall -- and a compression region on the vapor side of the liquid-vapor
interface. The results presented here raise the prospect that the correct
incorporation of fluid layering and the competition of compressive versus
shear effects at the contact line are the missing pieces to understanding the
nanoscale effects governing contact line motion.

\section{HDDFT}

For a simple fluid at the nanoscale, we assume that the stress tensor is
isotropic and that it is a function of the local density of the fluid. We
also assume that particle-particle interactions act non-locally as a body
force on other parts of the fluid. This is equivalent to replacing the
pressure-term in the momentum equation of the Navier-Stokes equations by $-
\nDensity \Grad \chemPot$, where the chemical potential $\chemPot$ is given
by
\begin{equation}
\chemPot = \frac{\delta \FE}{\delta \nDensity} + \Vext,
\end{equation}
where $\FE$ is the DFT free-energy functional, $\Vext$ is the estebral
potential and $n = \rho/m$ is the number density with $m$ the fluid particle
mass and $\rho$ the density. In contrast to equilibrium DFT, $\chemPot$ is
not a constant, but a function of space and time. This approach is in some
sense similar to the one followed diffuse interface models, which employ a
square-gradient expansion of the free energy of the system, and for which the
term $- \nDensity \Grad \chemPot$ accounts for surface tension effects via
the so-called Korteweg stress tensor~\cite{Sibley:2013:Unifying,David2013}.
However, the local expansion employed in diffuse-interface models has certain
shortcomings as highlighted in our previous effort in
Ref.~\cite{Antonio2012}. It does not capture layering effects and can lead to
multi-valued curves for the adsorption isotherm (effectively the thickness of
a film adsorbed on a flat substrate as a function of the distance of the
chemical potential from its saturation value), thus predicting phase
transitions in regions where there are none. There is also another more
subtle issue with diffuse-interface models. The models contain unknown and
ad-hoc parameters, in particular the interface width, say $\varepsilon$,
which is not known a priori. For sure, it has to be much larger than the
molecular scale, the hard sphere diameter $\sigma$, $\varepsilon \gg \sigma$,
as the diffuse-interface models do not have any molecular information and
this separation of scales is necessary. At the same time, $\varepsilon$ must
be much smaller than a characteristic macroscale in the system, say $l_{\rm
macro}$, $\varepsilon \ll l_{\rm macro}$, and in fact the ratio of the two is
taken to be small to show that the usual hydrodynamic equations for the
particular fluid flow setting at hand are recovered, e.g.~\cite{Marc2012}.

Here, we go beyond the local expansion and employ the DFT free-energy
functional, which is fully nonlocal. The interface width is no longer an
issue as the density gradient across the interface can be computed directly
(from which an effective interface width can be extracted, if so desired).
The fluid itself is assumed to be a Lennard-Jones (LJ) one, the standard
prototypical model of MD, with an intermolecular potential originating (at
least in part) from quantum mechanics, and appropriate for an idealised
system like liquid Argon (more involved models are often invoked in MD but
the results obtained with LJ are qualitatively satisfactory). The solid is
also taken as LJ and is modelled via an effective potential derived from
fluid-solid interparticle interactions. This mean-field approximation makes
the wall homogeneous and ideally-smooth with the same structure throughout,
thus avoiding details about its atomistic structure; details on DFT, and on
the fluid and solid models employed are given in Appendix~\ref{sec:DFT}.

The HDDFT model consists of the continuity and the momentum equations,
\begin{subequations}
\begin{align}
\diff{\nDensity}{t} + \Div \klamm{\nDensity \vel} &= 0,\\
\particleMass \nDensity \klamm{ \diff{\vel}{t} +\vel \cdot  \Grad \vel} &=
- \nDensity \Grad \klamm{\frac{\delta \FE}{\delta \nDensity}} + \Div \IrrevStressTensor, \label{eq:MomentumEq:General}
\end{align}
\end{subequations}
where $\IrrevStressTensor = \bulkViscosity \klamm{\Div \vel} +
\shearViscosity\klamm{  \klamm{\Grad \vel + (\Grad \vel)^T} - \frac{2}{3}
(\Div \vel)\IdMatrix }$ is the stress tensor, $\vel$ is the fluid velocity
and $\shearViscosity$ and $\bulkViscosity$ represent the shear (or dynamic)
and bulk (accounting for compressible effects) viscosities, respectively. For
simplicity, $\shearViscosity$ and $\bulkViscosity$ are assumed to depend
linearly on the local fluid density. Further details on the HDDFT model are
given in Appendix~\ref{sec:col-mol DDFT}, while values of liquid and vapour
viscosities for all computations throughout are given in
Table~\ref{tab:ViscosityData} there. Let us stress here that the Newtonian
law is not stipulated but can be derived as part of the HDDFT framework
\citep{Goddard:2013Unification}.

The model developed here assumes constant temperature throughout the fluid.
As discussed in Appendix~\ref{sec:DFT}, in the static case, the fluid
layering in the vicinity of a substrate computed by DFT is consistent with
Monte Carlo computations at the respective temperature. The dynamic
situation, however, is different: MD simulations employ thermostats to which
the nanoscale dynamics in highly inhomogeneous systems is
sensitive~\cite{Antonio2019}. The method proposed here fulfills the bulk
limit away from the wall commonly used to validate nanoscale models
\cite{smith2018moving}. But its cornerstone is the DFT free-energy functional
which as pointed out in Appendix~\ref{sec:DFT} is formulated in a constant
temperature -- constant chemical potential ensemble. As such the framework
proposed here is by design isothermal. This means that local temperature
changes due to viscous heat production are neglected. In principle, these can
be included by developing extended DDFTs, which also couple to an energy
equation, but the temperature changes must be slow (quasistatic). An
additional equation would inevitably increase the dimensionality of the
problem and therefore the computational cost, and would be against our
reductionist approach to formulate the simplest possible model. In
Appendix~\ref{sec:tempchange} we show that the total temperature increase is
likely not to exceed more than a few percent over the timescale of interest.

In our computations, we will consider chemical-potential-driven contact line
motion. This means that the contact angle is initially off equilibrium and
then approaches its equilibrium state. The bulk phases and fluid interfaces
are at equilibrium away from the contact line and motion is therefore driven
by nonequilibrium effects in the vicinity of the contact line alone. The
model does not require slip to resolve the contact line singularity. In fact
no slip is applied at the wall, which avoids the introduction of a slip
length as an empirical parameter, in agreement with the reductionist paradigm
advocated in this work.

\section{Fluid structure in the vicinity of the contact line}

\begin{figure}[t]
\centering
\hspace{-2cm} \includegraphics[width=10cm]{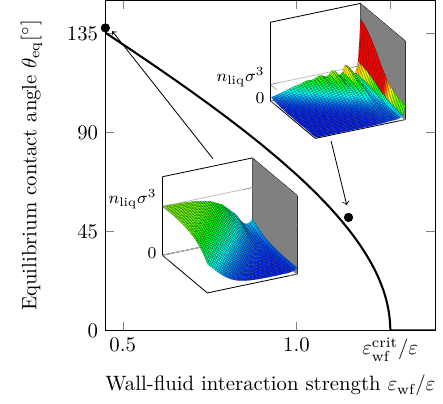}
\caption{Equilibrium contact angle $\thYoung$, computed via Young's equation based on the balance of wall-fluid and fluid-fluid surface tensions
(solid line), and two-dimensional DFT computations (filled circles)
as a function of normalized wall-fluid attraction $\LJWdepth/\varepsilon$ at the low temperature $k_B T = 0.75
\varepsilon$.
A comparison across the whole range of contact angles is given in
Refs.~\cite{Nold:FluidStructure:2014,nold2015nanoscale}.
Insets show three-dimensional profiles of the equilibrium number density $n$ on a $7\sigma
\times 7\sigma$ area close to the wall, shown in grey ($\sigma$ is the hard
sphere diameter). $\LJWdepthWetting$ marks the transition to complete
wetting.\label{fig:EqContactAngle}}
\end{figure}

The emphasis is on the fluid structure close to the contact line. At
equilibrium, the DFT computations show a well-defined defined interface in
the form of a wedge in contact with the substrate and with a well-defined
contact angle, as in our previous studies (which focused at
equilibrium)~\cite{Antonio2012,Nold:FluidStructure:2014,nold2015nanoscale,nold2017pseudospectral,
Nold2018} (e.g. Fig.~8 in Ref.~\cite{Antonio2012}). This wedge seems to
persist for distances sufficiently far from the substrate, and it should
eventually enter the macroscale, and should allow us to connect the micro-
with the macroscale. As illustrated in Fig.~\ref{fig:EqContactAngle}, the DFT
computations of the contact angle also follow closely Young's equation, again
in agreement with our previous
studies~\cite{Nold:FluidStructure:2014,nold2015nanoscale}, revealing also an
incredible fact about the equation: it is based on purely macroscopic
arguments, yet it is valid all the way to the nanoscale. As expected, the
contact angle decreases with increasing wall-fluid attraction
$\LJWdepth$~\cite{Antonio2012}. Strong wall-fluid attractions $\LJWdepth$
induce packing of fluid particles and large oscillations of the density
profile in the vicinity of the wall. The opposite effect is seen for large
contact angles and low wall-fluid attractions, where the formation of a
quasi-vapor film is
observed~\cite{Nold:FluidStructure:2014,nold2015nanoscale}. The connection
with disjoining pressure and interfacial Hamiltonian approaches was discussed
in Ref.~\cite{Nold:FluidStructure:2014} while in Ref.~\cite{Nold2018} another
incredible fact was revealed: density profiles at a three-phase contact line
may be described approximately by the density profile of a flat film provided
by the distance measured along the normal to the interface. As was already
alluded to in Ref.~\cite{Antonio2012}, to understand contact lines it is
essential that one understands the case of flat film on a substrate first.

\begin{figure}[t]
\centering
\hspace{-1cm} \includegraphics[width=14cm]{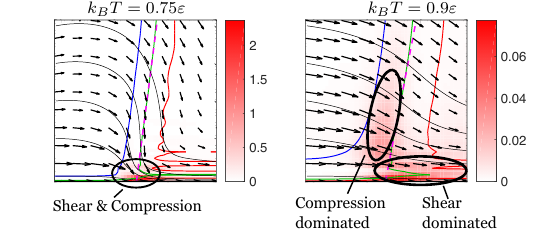}
\caption{HDDFT results for an
advancing contact line at two temperatures relaxing from $\thInitial = 90^\circ$ to $\thYoung = 60^\circ$ at $t=250 \tau$ on a $15\sigma
\times 15\sigma$ square domain next to the wall. For low temperatures, shear and compression dominated regions are located in the immediate vicinity of the contact line.
For high temperatures, the two regions separate in space,
and extend along the fluid-wall and fluid-fluid interfaces, respectively. Isolines (colored solid lines), velocity field
(arrows) and streamlines (black solid lines) are given relative to the motion
of the contact line. The blue, green and red isolines for the density
represent values $\klamm{\nDensity - \nDensityV}/\klamm{\nDensityL -
\nDensityV} = \{0.05,0.5,0.95\}$, respectively. The dashed magenta line, obtained as a
linear extrapolation of the green density isoline to the substrate, represents
the dynamic fluid interface and acts as a guide for the eye.
The blue and red isolines then are for the
vapour and liquid phases, respectively, in the immediate vicinity of the interface.
The colormap represents the entropy production $\viscousHeatProductionRescaled =  \viscousHeatProduction
\LJdiam^4/(\CLVel^2\sqrt{\varepsilon \particleMass})$, rescaled with the squared contact
line velocity $\CLVel$ (see Eq.~\ref{eq:ViscousHeatProduction}).
A detailed overview of the dynamic evolution of receding and advancing contact lines
at $k_B T = 0.75 \varepsilon$ and $0.9\varepsilon$ is given in Appendix~\ref{sec:dyn-evol-FlowPro}.
Separate plots of shear and compression for the advancing and receding case are given in
Fig.~\ref{fig:shearCompressionReceding}. Finally, he way an effective liquid-vapour interface
is calculated is given in the caption of Fig.~\ref{fig:Overview:kBT075} of Appendix~\ref{sec:dyn-evol-FlowPro};
the contact line position and contact angle have to be calculated with care -- the contact angle,
in particular, requires a careful measurement given all the terraces in the density profiles; details for the
calculation of contact line position--contact angle are given
in Appendix~\ref{compdetails}.
\label{fig:Fig3_SimulationsMovingCA}}
\end{figure}

Figure~\ref{fig:Fig3_SimulationsMovingCA} depicts velocity and density fields
for an advancing contact line. The density oscillations and fluid structure
observed at equilibrium persist in the dynamic state. Energy is dissipated
through transport of matter across the fluid interface as well as shear.
Classical thermodynamics quantifies the rate at which configurational and
kinetic energy is transformed into thermal energy as the viscous energy
dissipation $\viscousHeatProduction$ \cite{Lebon.2008}:
\begin{equation}
\viscousHeatProduction = \bulkViscosity \underbrace{\klamm{\Div \vel}^2}_{\textnormal{\parbox{1.2cm}{\centering compression}}} + \shearViscosity
\underbrace{\frac{1}{2} \klamm{(\Grad \vel + \klamm{\Grad \vel}^T) - \frac{2}{3}\klamm{\Div \vel}\IdMatrix}^2}_{\textnormal{shear}},
\label{eq:ViscousHeatProduction}
\end{equation}
where the square of the tensor is defined as the sum of its squared elements,
including a third dimension with all pertinent quantities set to zero. Away
from the contact line, the system moves in a purely convective manner.
However, sufficiently close to the contact line mass transport across the
liquid-vapor interface takes over, as seen by the compression of two
streamlines across the interface. As a consequence, a caterpillar-rolling
motion--like behavior is observed in the liquid phase, whilst the vapor phase
is almost entirely compressed into, or decompressed from, the liquid phase
for advancing and receding contact lines, respectively.

In Appendix~\ref{sec:tempchange} we demonstrate that the total temperature
increase due to viscous energy dissipation is likely not to exceed more than
a few percent over the timescale of interest. This is due to the
(realistically) low contact line velocities (Figs~\ref{fig:Overview:kBT075}
and~\ref{fig:Overview:kBT09} of Appendix~\ref{sec:dyn-evol-FlowPro}) together
with the particular spatial organisation of the dissipation regime.

\section{Emergent contact line behavior at the nanoscale}

We will be looking at two temperatures: the low temperature $k_B T =
0.75\varepsilon$, and the high temperature $k_B T = 0.9\varepsilon$,
relatively close to the critical point. Of course, DFT and consequently
HDDFT, are mean-field approximations which ignore fluctuations and therefore
are less reliable as criticality is approached, where fluctuations may play a
dominant role. It is for this reason exactly that we avoid being too close to
criticality. Our overarching objective here is to expose certain trends in
contact line motion and uncover new patterns and insights as the temperature
increases. By taking the temperature relatively close to the critical point a
certain qualitative picture emerges, and our results show a clear pattern as
detailed below.

\begin{figure}[htp]
\centering
\includegraphics[width=0.8\linewidth]{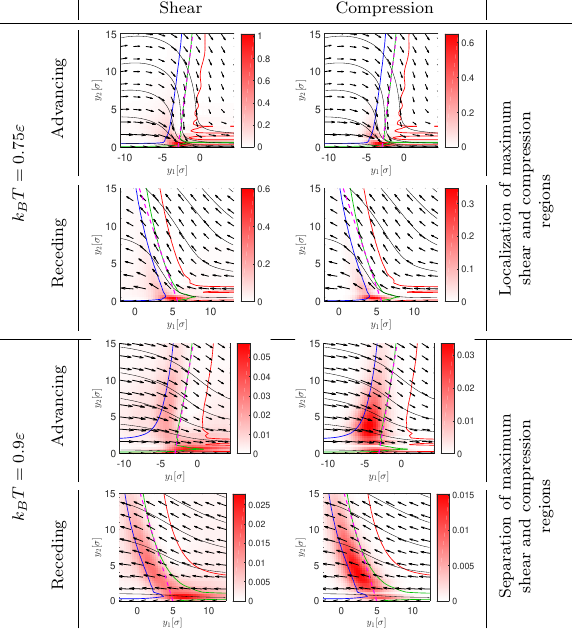}	
\caption{When increasing the temperature, here from $k_BT=0.75\varepsilon$ to $0.9\varepsilon$, two regions emerge:
a compressive region which favors the vapor side of the liquid-vapor interface, and a shear region
in the vicinity of the wall. The plot represents the compressive and shear contributions to the viscous energy
dissipation (see braces in Eq.~\ref{eq:ViscousHeatProduction}) on a $15\sigma \times 15 \sigma$ square domain
next to the wall, normalized with $\CLVel^2$, in units of $1/\sigma^2$  for an advancing contact line
(parameters as in Fig.~\ref{fig:Fig3_SimulationsMovingCA}).
The blue, green and red isolines for the density represent values $\klamm{\nDensity - \nDensityV}/\klamm{\nDensityL -
\nDensityV} = \{0.05,0.5,0.95\}$, respectively. For advancing contact lines, an effective slip region extends
next to the wall into the liquid side (slip is discussed in Appendix~\ref{sliplength}). At high temperatures, the peak of the
compression distribution is moved away from the wall. Velocity fields and streamlines are given relative to the moving contact line.
For receding contact lines, the peak of the distributions is shifted to the
vapor side.
\label{fig:shearCompressionReceding}}
\end{figure}

\begin{figure}
\centering	
\includegraphics[width=12cm]{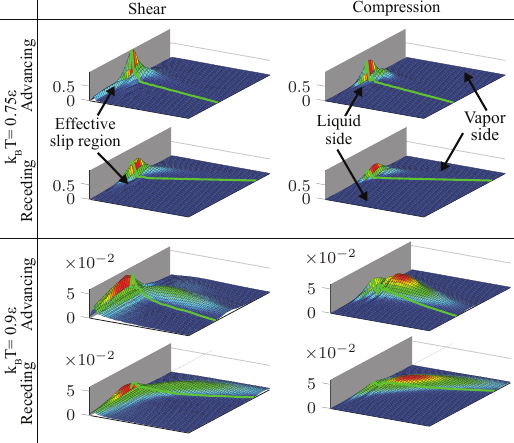}		
	\caption{Data as in Fig. \ref{fig:shearCompressionReceding}, re-plotted here to emphasize magnitude of distribution in space.
    \label{fig:shearCompressionReceding:2}}
\end{figure}

At temperature $k_B T = 0.75\varepsilon$, most of the transfer of mechanical
into thermal energy is concentrated in the immediate vicinity of the contact
line. In contrast, raising the temperature to $k_B T = 0.9\varepsilon$
removes the localization and viscous energy dissipation spreads along the
wall and the vapor side of the fluid interface, as shown in
Figs~\ref{fig:Fig3_SimulationsMovingCA} and
~\ref{fig:shearCompressionReceding}, with the dissipative region being
separated into compressive and shear regions. Whilst some change of spatial
distribution with temperature is expected, the difference exceeds the change
suggested by the increase of the liquid-vapor interface width with
temperature from $4.2\sigma$ for $k_BT = 0.75\varepsilon$ to $7.3\sigma$ for
$k_BT = 0.9\varepsilon$.
A separate study of compressive and shear regions reveals a distinct spatial
pattern as illustrated in Figs~\ref{fig:shearCompressionReceding} and
~\ref{fig:shearCompressionReceding:2}: (a) A layer of high shear extending
along the wall in the liquid side of the fluid. This layer is more pronounced
in the advancing than the receding case; (b) A compression region extending
along the vapor side of the interface. Furthermore, shear and compression are
of the same order of magnitude, and exhibit a similar distribution, albeit
with peaks located in different regions. In fact, in the receding case, the
peak of energy dissipation in the vicinity of the contact line is slightly
moved to the energetically more favorable vapor side.

The natural question then is whether shear and compression effects control
contact line motion at the same rate. To address this, we successively
lowered and raised the shear and bulk (or compressive) viscosities,
$\shearViscosity$ and $\bulkViscosity$, respectively, as shown in
Table~\ref{tab:ViscosityVariation} of Appendix~\ref{viscochange}. As
highlighted in the Appendix, contact line motion is more sensitive to changes
of liquid than vapor shear viscosity. Interestingly, the opposite effect is
seen for the bulk (or compressive viscosity) and now contact line motion is
less sensitive to changes of liquid than vapor bulk viscosity. This is
consistent with the locations of the high shear-layer and high compression
regions, in the liquid and the vapor side of the contact line, respectively.
The results also reveal that the bulk viscosity has a disproportionally high
impact on the velocity of advancing contact lines at low temperatures.

Finally, like with the equilibrium case, the configuration for the
liquid-vapour interface observed in Fig.~\ref{fig:shearCompressionReceding}
seems to persist for distances sufficiently far from the wall (the maximum
value of $y_2$ depicted in the figure is 15 molecular diameters), and so it
should enter the macroscale allowing us to connect the micro- with the
macroscale for moving contact lines.

\section{Emergent contact line behavior at the mesoscale\label{emergent-meso}}

Shear effects play an important role in controlling contact line motion in
the vicinity of the contact line~\cite{McGraw1168}. The thickness of the
shear layer is of the order of a particle diameter and spreads along the wall
into the liquid and vapor phases as the temperature is increased (see
Fig.~\ref{fig:shearCompressionReceding}, and
Figs.~\ref{fig:FlowPropertiesAroundTheCL},
\ref{fig:FlowPropertiesAroundTheCL:09} in
Appendix~\ref{sec:dyn-evol-FlowPro}). It suggests the existence of an
effective slip region interpolating between the no-slip region/wall and the
bulk fluid. The width of this region is restricted by the oscillatory density
structure induced by particle stratification close to the wall. Inferring the
slip length from the velocity profile perpendicular to the wall reveals that,
contrary to popular belief in fluid mechanics, it is not a robust parameter
with respect to the choice of effective shear-layer thickness, precisely
because of the fluid density oscillations in the vicinity of the well (see
Appendix~\ref{sliplength} and Fig.~\ref{fig:SlipVelocity} there). We thus
infer that, whilst the first layer of fluid particles acts as a
slip-facilitating layer, an effective slip length should be seen at best as
an abstract coarse-grained parameter, of the order of the particle diameter,
which includes more effects than just physical slip at an idealized
zero-thickness shear layer.

Conversely, the contact line friction coefficient $\frictionCoefficientMKT$
turns out to be a more robust measure than the slip length as illustrated in
Figs.~\ref{fig:MKT:Robustness:adv:kbT075}--\ref{fig:MKT:Robustness:rec:kbT09})
of Appendix~\ref{contactlinefriccomp}. It is precisely for this reason that
it is used a benchmark for the accuracy of our numerical scheme in
Appendix~\ref{compdetailsIt}.It represents a relatively simple way to relate
the contact angle $\theta$ with the contact line velocity $\CLVel$:
%
\begin{equation}
\frictionCoefficientMKT \CLVel =  \surfaceTensionLV \klamm{ \cos \theta - \cos \thYoung},  \label{eq:Nano:DDFT:MKTmodel}
\end{equation}
and, like Young's equation, it is valid all the way to the nanoscale even
though it was obtained from macroscopic arguments.

\begin{figure}[h!tp]
\centering
\includegraphics[width=\linewidth]{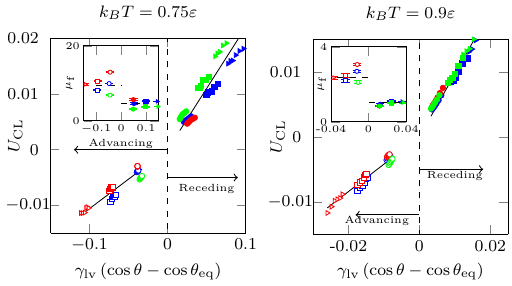}
\caption{\label{fig:Nano:DDFT:CheckMKT}
{Contact line velocity, $\CLVel$, in multiples of $\sigma\tau^{-1}$ vs. force, $\surfaceTensionLV \klamm{ \cos \theta - \cos \thYoung}$,
acting on the contact line in multiples of
$\varepsilon/\sigma^2$ at times $t = 125\ldots  250 \tau$ and $200\ldots 400\tau$. The black solid lines are fits to the prediction of
Eq.~\ref{eq:Nano:DDFT:MKTmodel} with resulting friction coefficients
$\frictionCoefficientMKT(k_BT = 0.75 \varepsilon) = (9.5,4.5)$, $\frictionCoefficientMKT(k_BT = 0.9 \varepsilon) = (2.4,1.0)$ in multiples of $\sqrt{\varepsilon m}/\sigma^2$
for the advancing and the receding case, respectively.
Simulations with equilibrium contact angle $\thYoung$ of $70^\circ, 90^\circ$ and $110^\circ$ are represented by red, blue and green symbols, respectively.
Circles, squares and triangles represent values with $|\thYoung-\thInitial | = 10^\circ,20^\circ$ and $30^\circ$, respectively.
Filled and empty symbols represent receding and advancing contact lines, respectively.
Insets display the contact line friction as a function of $\surfaceTensionLV (\cos \thInitial - \cos \thYoung)$, in multiples of $\varepsilon/\sigma^2$.}}
\end{figure}

The constitutive relation between the friction force per unit length and the
off-equilibrium driving force can be constrained using MD simulations: Ren
{\it et al.} \cite{ren2010continuum} showed that nonlinear effects can indeed
be neglected up to velocities of $\approx 1 \sigma\tau^{-1}$, i.e. $50$ times
larger that the highest contact line velocities computed here.
Figure~\ref{fig:Nano:DDFT:CheckMKT} illustrates that
Eq.~\ref{eq:Nano:DDFT:MKTmodel} fits the (highly nonlinear and nonlocal)
HDDFT well, both as the contact line equilibrates over time, and for varying
initial conditions.

\begin{figure}[h!tp]
\centering
\includegraphics[width=\linewidth]{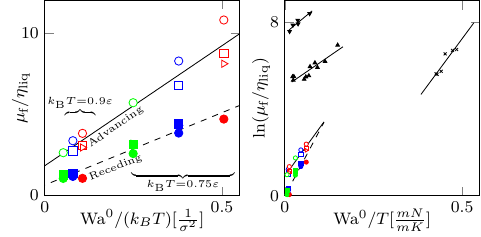}
\vspace{-0.2cm}
\caption{Contact line friction $\frictionCoefficientMKT$ varies with $\shearViscosityL$ and $\Wa/(k_BT)$.
Left: Simulations towards equilibrium contact angles $\thYoung$ of $70^\circ, 90^\circ$ and $110^\circ$ are represented by red, blue and green symbols, respectively.
Circles, squares and triangles represent values with $|\thYoung-\thInitial | = 10^\circ,20^\circ$ and $30^\circ$, respectively.
Filled and empty symbols represent receding and advancing contact lines, respectively.
Right: Black symbols are experimental data for spreading
aqueous glycerol drops on glass ($\times$, \cite{duvivier2011experimental}), a
moving droplet of dodecane in water ($\blacktriangle$, \cite{ramiasa2011contact})
and an air bubble in water ($\blacktriangledown$, \cite{ramiasa2011contact}). For
the data from \cite{ramiasa2011contact}, the sum of both fluid viscosities
was used to rescale the contact line friction. Here, we have employed a typical value for the hard sphere diameter that of Argon, $\sigma = 3.3952 \AA$,
as a characteristic length scale for our data~\cite{vrabec2001set}. Black lines are linear fits to the data.
\label{fig:5}
}
\end{figure}

The contact line friction $\frictionCoefficientMKT$ depends on temperature,
the wall-liquid interaction strength, and shear viscosity. We now compare our
results with an extension of the MKT model by Blake and De Coninck
\citep{blake2002influence}, based on activation energies of single particle
displacements, to include viscous effects:
\begin{equation}
\frictionCoefficientMKT = \klamm{\frac{v_L}{\bar\lambda^3}} \shearViscosityL \exp\klamm{\frac{\bar\lambda^2 \Wa}{k_BT} }, \label{eq:MKT:CLFriction:BlakeDeConninck}
\end{equation}
where $\Wa = \surfaceTensionLV\klamm{1 + \cos \thYoung}$ is the reversible
work of adhesion, $\bar\lambda$ (also known as the Young-Dupr\'e
equation~\cite{Schrader1995}) is the molecular displacement length, and $v_L$
is the ``volume of the unit of flow". Here, the variations in the argument
are small, and Eq. \ref{eq:MKT:CLFriction:BlakeDeConninck} can be linearized.
Figure~\ref{fig:5} reveals a very good agreement across temperatures and
substrate strengths (the effect of changing the shear viscosity is
illustrated in Fig.~\ref{fig:ChangeOfContactLineFrictionWithViscosity} of
Appendix~\ref{contactlinefriccomp}). This linear relationship changes with
the direction of flow: for both low and high temperatures, the contact line
friction is approximately $50\%$ lower for the receding compared to the
advancing case, consistent with our computational result that dewetting is
faster than wetting, as can be inferred from
Fig.~\ref{fig:Nano:DDFT:CheckMKT}. The trends reported here also compare well
with experimental data \citep{duvivier2011experimental,ramiasa2011contact},
albeit with a different prefactor (see Fig.~\ref{fig:5}, right panel,
Fig.~\ref{fig:contactLineFriction_ComparisonWithExperiment} of
Appendix~\ref{contactlinefriccomp} for a comparison with MD data).
Unfortunately the range of data is not sufficient to infer straight lines
with full certainty. Let us also highlight that unlike MKT where the contact
line friction is a free parameter, HDDFT enables to go a step further than
MKT and benchmark the friction coefficient as a function of viscosity and the
system's fundamental parameters, i.e. the strength of fluid-fluid and
wall-fluid interactions, and temperature.
%

\section{Closing remarks and discussion}

We have coupled a particle-based approach, namely DDFT, with continuum
hydrodynamics, to study an isothermal simple-fluid liquid-vapour contact line
at low and high temperatures. The resulting multiscale continuum framework,
referred to as HDDFT, enables us to probe the nanoscale behavior of the
moving contact line for a wide range of temperatures, including values
relatively close to the critical point, a regime, which especially for
liquid-vapour systems is notoriously difficult to capture. HDDFT avoids the
limitations of other approaches, e.g. the small signal-to-noise ratios
present in MD, but also experimental techniques, whilst still retaining
information from the nanoscale through DFT.

\begin{figure}
\centering
\includegraphics[width=\linewidth]{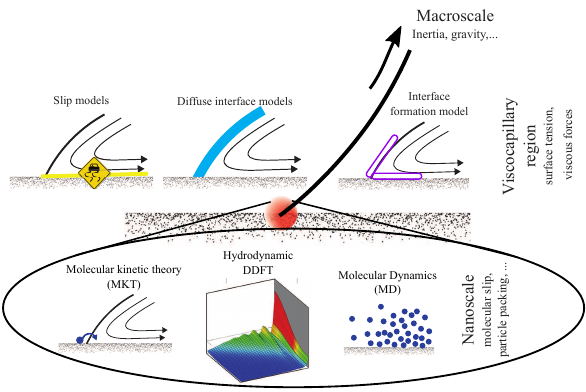}
\caption{Schematic of contact line regions and corresponding approaches and models. The viscocapillary region, discussed in the Introduction,
is defined by a balance of surface tension and viscous forces.
Slip models~\cite{Voinov:1976fk}, diffuse interface models~\cite{Seppecher:1996vn,anderson:2001phase,Sibley:2013:Unifying} and the interface formation model
\cite{Shikhmurzaev:1993uq} resolve this regime.
These models include at least one free parameter which captures the scale separation between the micro- and the macroscopic region. The
behavior at the nanoscale is governed by various factors including microscopic slip, interfaces with finite thickness and surface layers in the region near
to an interface between the liquid and the substrate-vapour.
MKT \cite{Blake.19697,blake2002influence} resolves this region by postulating an activation energy for molecular displacement.
MD resolves individual particles in space and time.
The HDDFT model introduced in this work includes fluid layering and compressive and shear effects.
\label{fig:OverviewConclusion}}
\end{figure}

Our computations with HDDFT suggest an intriguing length scale separation for
large temperatures: energy dissipation due to shear effects is mostly
concentrated at distances of a few particle diameters from the wall.
Dissipation due to compressive effects, however, exhibits a peak which
extends into the vapour side of the fluid interface. Both shear and
compression have a distinct impact on contact line motion which depends on
direction and temperature. The width of the shear layer and associated
effective slip region is controlled by the hard sphere particle diameter. In
contrast, the compressive region is controlled by a temperature-dependent
length scale. Interestingly, this compressive region is reminiscent of the
interface layer stipulated by Shikhmurzaev in the context of the so-called
interface formation model \cite{Shikhmurzaev:1993uq} -- a detailed
examination of the interface formation model was provided in our previous
work in Ref.~\cite{David2012}. Its effect in fact is not negligible, as even
in the limit of zero liquid shear viscosity, the contact line friction is
approximately $50\%$ of its original value (see
Fig.~\ref{fig:ChangeOfContactLineFrictionWithViscosity} of
Appendix~\ref{contactlinefriccomp}). This means that in addition to viscous
effects in the bulk, slip effects at the wall, and dissipation in the
thin-film regime, such as outlined by Ren {\it et al.}
\cite{ren2010continuum}, the relative contribution of dissipation due to
compressive effects needs to be taken into account and may offset contact
line friction. We note, however, that in the spirit of the reductionist
paradigm followed here, our model does not include a slip length and imposes
instead the no-slip condition at the wall. This introduces a strong
dissipative mechanism, in contrast to slip models or models of molecular
displacement/MKT. Nevertheless, we can estimate that the dissipation-driven
heat produced by the dynamics assuming isothermal conditions would lead to a
temperature increase that would not exceed $k_BT/\varepsilon \approx 0.06$
(Appendix~\ref{sec:tempchange}). A schematic summarising the contact line
regions and different approaches and models for contact line motion (slip,
diffuse interfaces, interface formation model, MKT, MD and HDDFT in the
present study), can be found in Fig.~\ref{fig:OverviewConclusion}. The
asymptotics of diffuse interface models for the moving contact line problem
were analysed in detail in our previous works in
Refs~\cite{sibley2013moving,David2013}.

Our study appears to corroborate
experimental~\cite{duvivier2011experimental,ramiasa2011contact,zhao2017analyzing}
and MD-based evidence~\cite{lukyanov2016dynamic} that the main source of
dissipation stems from the deviation of the dynamic contact angle from its
equilibrium value. HDDFT is also in line with the contact-line friction model
(cf.~Eqs.~\ref{eq:Nano:DDFT:MKTmodel},
\ref{eq:MKT:CLFriction:BlakeDeConninck}), often referred to as (linear) MKT
\cite{Blake.19697} -- which starts from completely different model
assumptions. This provides further evidence that the contact line friction,
$\frictionCoefficientMKT$, acts as an abstract measure, similar to the slip
length in the viscocapillary regime, reflecting the interplay of different
physical effects at the nanoscale. However, extracting
$\frictionCoefficientMKT$ from the nanoscale is more robust than doing so for
the slip length. In fact our computations show a generalised linear
relationship between $\frictionCoefficientMKT$ and $\klamm{ \cos \theta -
\cos \thYoung}$ across substrate strengths and temperatures as well as fluid
viscosities, $\shearViscosityL$ (Figs~\ref{fig:Nano:DDFT:CheckMKT}
and~\ref{fig:ChangeOfContactLineFrictionWithViscosity} of
Appendix~\ref{contactlinefriccomp}). Quantitative comparison with
experimental data \cite{duvivier2011experimental,ramiasa2011contact} reveals
a similar rate of increase of
$\log(\frictionCoefficientMKT/\shearViscosityL)$ with $\textnormal{Wa}^0/T$
in Eq.~\ref{eq:MKT:CLFriction:BlakeDeConninck}, where $\textnormal{Wa}^0$ is
the work of adhesion, across different fluid-substrate pairs, but with a
different offset (Fig.~\ref{fig:5}, right panel). This difference may emerge
from the influence of volatility in the compressible region at the
liquid-vapor interface. Alternatively, layering effects in the effective slip
region close to the wall can affect contact line friction, such as observed
close to complete wetting \cite{wu2017capillary}. We note that no convex
nanobending of the fluid interface as reported in the experiments by Chen
{\it et al.} \cite{chen2014convex} is found here, but these authors
considered (molecular) highly viscous non-volatile liquids.

\section*{Acknowledgements}
We acknowledge financial support from Imperial College through a DTG
International Studentship, the Joachim Herz foundation through an ``Add-on
Fellowships for Interdisciplinary Science", the ERC-EPSRC Frontier Research
Guarantee via Grant No. EP/X038645, the ERC via Advanced Grant No. 247031,
and the EPSRC via Grant No. EP/L020564. We are grateful to the anonymous
Referees for valuable comments and critical suggestions. Special thanks to
Mr. Antonio Malpica-Morales from the Department of Chemical Engineering of
Imperial for his assistance with Figs 10-13 and 15-18.

\appendix

\section{DFT\label{sec:DFT}}

Equilibrium density distributions are computed employing
classical DFT, which describes an equilibrium many-body system uniquely
by a functional $\GrandPotential[\nDensity]$ ~\cite{Evans,Wu-DFT}, which is minimised with
respect to the number density distribution $\nDensity$. For a single-component system,
\begin{equation*}
\GrandPotential[\nDensity] = \FE[\nDensity] + \int \nDensity(\pos) \klammCurl{\Vext(\pos) - \chemPot} \dI\pos,
\end{equation*}
where at equilibrium $\FE$ is the intrinsic Helmholtz free-energy functional,
$\Vext$ is the external potential of the system and $\GrandPotential$ is the
grand potential. $\chemPot$ is the chemical potential, which tunes the system
towards the liquid or the vapor state, here always chosen to be at
equilibrium. We highlight that our DFT is formulated in a constant
temperature -- constant chemical potential ensemble.

The fluid is a simple LJ fluid with a potential with depth $\LJdepth$ and
length scale $\LJdiam$, and split $\FE[\nDensity]$ into an ideal-gas
contribution and contributions due to short-range hard-sphere exclusion
effects, $\FEhs$, and long-range attractive interactions, $\FEattr$. $\FEhs$
is approximated by Rosenfeld's fundamental measure theory
(FMT)~\cite{Rosenfeld:1989qc,Roth:2010fk}, with $\sigma$ used as the
hard-sphere diameter. Let us note here that FMT enables to capture the
near-wall fluid layering leading to oscillatory fluid density there, or
indeed layering in the bulk fluid at freezing conditions/low temperatures,
and, at equilibrium, it has been shown to be in excellent agreement with
Monte Carlo simulations~\cite{nold2017pseudospectral}. On the other hand, for
liquid-vapour systems, the so-called local-density approximation (LDA)
(e.g.~\citep{Antonio2012,Peter2016b}) is sufficient as it captures the main
features and is in qualitative agreement with more sophisticated
approximations.

The attractive particle-particle interactions are treated in a perturbative
mean-field manner~\cite{Zwanzig:1954cq,TangWu:2003,Lasse:etal:2013} as
\begin{equation*}
\FEattr[\nDensity] = \frac{1}{2 } \iint  \phi_{\text{attr}}({|\pos - \pos'|})
\nDensity(\pos)\nDensity(\pos') \dI\pos' \dI\pos,
\end{equation*}
 where
$\BHattr\klamm{r} = 4 \LJdepth\klamm{ \klamm{{\LJdiam}/{r}}^{12} -
\klamm{{\LJdiam}/{r}}^{6}} \Theta\klamm{r-\sigma}$ with radial cutoff at $r_c
= 2.5\LJdiam$.  The wall is also assumed to be LJ and $\Vext$ is obtained by
integrating the LJ wall-fluid particle interactions with depth $\LJWdepth$
over a homogeneous flat wall. The wall then acts as a uniform effective
boundary, thus sidestepping details about its atomistic makeup. For
simplicity, the wall particles are assumed to have the same diameter as the
fluid particles, $\sigma$. $\tau
\defi \LJdiam \sqrt{\frac{\particleMass}{\varepsilon}}$ is the time scale,
with $m$ the particle mass. The energy scale is $\varepsilon \defi -
\frac{9}{32\pi} \int_{\mathbb{R}^3} \BHattr\klamm{|\pos|} \dI \pos$, which
ensures that the bulk effect of the attractive contribution to the free
energy $\int_{\mathbb{R}^3} \BHattr\klamm{|\pos|} \dI \pos$ is independent of
choice of the radial cutoff parameter $r_c$.

\section{Colloidal and Molecular DDFT\label{sec:col-mol DDFT}}

Equations analogous to Eq.~\ref{eq:MomentumEq:General} but with damping term,
have been derived in a statistical mechanics of fluids framework for
colloidal fluids starting from the microscopic equations of motion of the
individual particles ~\cite{Archer:2009fk,Goddard:2013Unification}. However,
in contrast to colloidal fluids, for molecular or atomic fluids, no damping
effect exists. The system is fully determined by Newton's equations of motion
at the molecular scale. Formally, this simplifies the equations and the link
to the generalized momentum equation derived for colloidal systems can still
be made by following the derivation of the continuity and momentum equations
from Kramer's equation as outlined in
Refs.~\cite{Archer:2009fk,Goddard:2012general,Goddard:2013Unification} and
then setting the damping coefficient to zero. The main approximations that
allow us to make progress are: (i) local equilibrium and expansion around
local equilibrium; (ii) the adiabatic approximation, i.e. the higher-body
correlations in the non-equilibrium fluid are approximated by those of an
equilibrium fluid with the same density, and the corresponding sum rule holds
out of equilibrium (statistical mechanical sum rules are discussed
in~\cite{Henderson1992}). The zeroth and first moment of Kramer's equation
then yield:
\begin{subequations}
\begin{align}
\diff{\nDensity}{t} + \Div \klamm{\nDensity \vel} &= 0,\\
\particleMass \nDensity \klamm{ \diff{\vel}{t} +\vel \cdot  \Grad \vel}
&= - \nDensity \Grad \klamm{\frac{\delta \FE}{\delta \nDensity}} + \Div \IrrevStressTensor, \label{eq:MomentumEq:General:Appendix}
\end{align}
\end{subequations}
where $\IrrevStressTensor = \bulkViscosity \klamm{\Div \vel} +
\shearViscosity\klamm{  \klamm{\Grad \vel + (\Grad \vel)^T} - \frac{2}{3}
(\Div \vel)\IdMatrix }$ is the stress tensor, $\vel$ is the fluid velocity,
$\shearViscosity$ is the usual shear or dynamic viscosity and
$\bulkViscosity$ is the bulk viscosity accounting for compressible effects.
Unfortunately, at this level of approximation the above system remains
unclosed in that $\shearViscosity$ and $\bulkViscosity$ depend on the
first-order terms of the expansions of the one-body distribution and the
two-body density, and cannot be written in closed form as functions of the
local density or velocity. Here we follow~\cite{Sibley:2013} and employ
empirical constitutive laws for the bulk and the shear viscosities as linear
functions of density, interpolated between their liquid and vapour values,
which for the shear viscosity can approximate MD well under certain
conditions~\cite{Morciano:2017,Antonio2019}. Values for liquid and vapour
viscosities correspond to MD results close to the saturation densities at
temperatures $T/T_{\textnormal{crit}}$; see Table ~\ref{tab:ViscosityData}.

\begin{table}[htp]
\begin{center}
\addtolength{\tabcolsep}{5pt}
\begin{tabular}{ccc|c|cccc}
$k_BT/\varepsilon$ & $\nDensityV\LJdiam^3$ & $\nDensityL\LJdiam^3$ & $\surfaceTensionLV$ & $\shearViscosityV$ & $\shearViscosityL$ & $\bulkViscosityV$ & $\bulkViscosityL$\\
\hline
0.75 & 0.028 & 0.622 & 0.285 & 0.10 & 1.2 & 0 & 5.5\\
0.9 & 0.086 & 0.458 & 0.072 & 0.15 & 0.8 & 0 & 2.0
\end{tabular}
\addtolength{\tabcolsep}{-5pt}
\end{center}
\caption{Liquid and vapor shear and bulk viscosities for a simple fluid at different temperatures.
{\normalfont Bulk densities and the surface tension are obtained from DFT computations.
Critical temperature and densities from the DFT model are $k_BT_{\textnormal{crit}} /\varepsilon = 1.001$ and $\nDensityCrit \sigma^3 = 0.246$.
The liquid-vapor surface tension $\surfaceTensionLV$ is given in multiples of $\varepsilon/\sigma^2$ and viscosities are given as multiples of
$\sqrt{\varepsilon \particleMass}/\LJdiam^2$. Shear viscosities are obtained from Figs. 7 and 8 of Ref. \cite{Meier:2004transport}, and bulk viscosities from
Fig. 4 of Ref.~\cite{Meier:2005:BulkViscosities}, both for the corresponding values of $T/T_{\textnormal{crit}}$. It is noteworthy that the liquid viscosity data exhibits
large variability with density, see e.g. also data in Ref.~\cite{baidakov2014metastable}.
\label{tab:ViscosityData}}}
\end{table}%

It is noteworthy that the HDDFT model we propose here represents an
alternative approach to MD-based numerical methods
\cite{Hadjiconstantinou,smith2018moving} to probe the link between basic
nanoscale ingredients with the macroscale. But there are a few caveats and
simplifying assumptions. Coupling of DFT in a hydrodynamic setting means that
nonlocal correlations between the density profile and non-isotropic effects
in momentum transport are not taken into account. Instead, we assume scalars
for the viscosities (local linear functional dependence of the viscosities on
the density), and the stress tensor to be isotropic.
Mean-field models also do not account for fluctuations
\cite{MacDowellBenet:2014,Miguel2017}.
At temperatures close to the critical point, fluctuation-assisted nucleation,
which we neglect here, may play a role for contact line movement
\cite{Xu:2010kx}. We have extended the equations
in~\ref{eq:MomentumEq:General:Appendix} to include
fluctuations~\cite{Miguel2017}; this derivation is rigorous and supports the
intuitive treatment of the original theory by Landau and Lifshitz of
stochastic Navier-Stokes~\cite{Landau1980}. Accounting for these effects in
contact line motion is not within the scope of the present study and needs to
be tackled in future dedicated theoretical and computational work.

\section{Analysis of temperature variation due to energy
dissipation\label{sec:tempchange}}

The energy dissipated by shear and compression will heat up the fluid leading
to local temperature changes. Here we assess these temperature changes.
Analogous to Fig.~\ref{fig:Fig3_SimulationsMovingCA}, we employ molar heat
capacity for gaseous \cite{chase1998j} and liquid \cite{gladun1971specific}
Argon as a model system:
\begin{align}
c_m = 18\ldots 21 \frac{J}{\textnormal{mol} \cdot K} \approx 20 \frac{J}{\textnormal{mol} \cdot K}.
\end{align}
For small perturbations, the local rate of temperature change can be
approximated as:
\begin{align}
\frac{k_B}{\varepsilon} \Delta \dot{T} &= \frac{k_B}{\varepsilon} \frac{\viscousHeatProduction}{c_m n}\\
&\approx \frac{0.41}{\varepsilon} \frac{\viscousHeatProduction}{n} \label{eq:deltaT:Approx1}.
\end{align}

Let us now compute an upper estimate for the temperature rise, by considering
an advancing contact angle at $k_BT = 0.75 \varepsilon$, which yields large
levels of concentrated compression and shear in the vicinity of the contact
line, as illustrated in Fig. \ref{fig:shearCompressionReceding}. The contact
line velocity for this case is approximately $\CLVel = -0.01 \sigma/\tau$
(see Fig.~\ref{fig:Overview:kBT075}). Employing the larger values for liquid
shear and bulk viscosities (see Table~\ref{tab:ViscosityData}) and the maxima
of shear and compression shown in Fig.~\ref{fig:shearCompressionReceding}, we
obtain that the viscous heat production in Eq.~\ref{eq:ViscousHeatProduction}
and can be written as
\begin{align}
\viscousHeatProduction &=
\CLVel^2
\klammCurl{
\bulkViscosity \klamm{\frac{\klamm{\Div \vel}^2}{\CLVel^2} }
+
\shearViscosity\frac{\klamm{(\Grad \vel + \klamm{\Grad \vel}^T) - \frac{2}{3}\klamm{\Div \vel}\IdMatrix}^2}{2\CLVel^2}
}\\
&\approx \klamm{0.01 \frac{\sigma}{\tau}}^2 \klammCurl{1.2 \frac{\sqrt{\varepsilon m}}{\sigma^2} \cdot 1\frac{1}{\sigma^2} + 5.5 \frac{\sqrt{\varepsilon m}}{\sigma^2} \cdot 0.6\frac{1}{\sigma^2} } \\
&\approx 4.5 \cdot 10^{-4} \frac{\varepsilon}{\tau \sigma^3},
\end{align}
where we used that $\tau = \sigma \sqrt{m/\varepsilon}$. Adopting an estimate
for the density of $0.622 \frac{1}{\sigma^3}$ (see Table
\ref{tab:ViscosityData}), leads after insertion into
Eq.~\ref{eq:deltaT:Approx1} to
\begin{align}
\frac{k_B}{\varepsilon} \Delta \dot{T} \approx 3 \cdot 10^{-4} \frac{1}{\tau}.
\end{align}
Here we study quasi-steady contact line movement in a time window of between
approximately $200\tau$ and $400\tau$ (see Fig.~\ref{fig:Overview:kBT075}).
Therefore, the total temperature change within this time window for the
vicinity of the contact line does not exceed
\begin{align}
\frac{k_B}{\varepsilon} \Delta T \approx 0.06,
\end{align}
leading to an increase from $0.75$ to $0.81$ for ${k_B T}/{\varepsilon}$. The
true temperature increase is likely to be lower, as heat flux to areas
neighboring the contact line and adjacent wall will likely reduce the
localised heating effect.

\section{Computational details\label{compdetails}}

\subsection{Numerical scheme}

The density profiles are obtained using a highly accurate and robust
pseudospectral scheme with $40$ grid points in each direction, as described
in our previous work in Ref.~\cite{nold2017pseudospectral}. In short, the
unit square $[-1,1]\times[-1,1]$ computational grid of Chebyshev polynomials
is conformally mapped onto to the semi-infinite physical space, which aligns
the grid with the initial contact angle. $50\%$ of the discretization points
are mapped onto $[-L_1,L_1]$ and $[0,L_2]$ in the $y_1$ and $y_2$ directions,
respectively. For $k_BT=0.75\varepsilon$, $L_1 = 4\sigma$ and $L_2 =
2\sigma$. For $k_BT =0.9\varepsilon$, $L_1 = 6\sigma$ and $L_2 = 2\sigma$.
The discretisation of the domain is skewed to increase the number of points
near the fluid interface where higher gradients are expected. The convolution
matrices needed for the FMT computations are also implemented using spectral
methods. For this, different mappings of the collocation points from the
computational to the physical domain are needed, depending on the proximity
of the respective point from the wall (collocation points in the wall must be
avoided).

To solve the dynamic equations, we employ the variable-step, variable-order
solver ode15s from Matlab.

\begin{figure}[h!tp]
\centering
\includegraphics{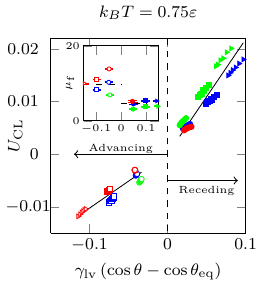}
\includegraphics{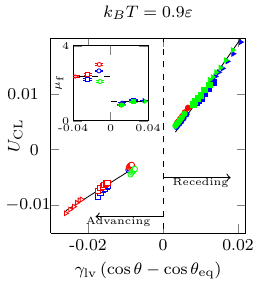}\\
\vspace{1cm}
\begin{tabular}{m{8cm}||cc|cc}
Mean contact line friction $\frictionCoefficientMKT$ [$\sqrt{\varepsilon m}/\sigma^2$]
&
\multicolumn{2}{c}{$k_BT=0.75\varepsilon$}& \multicolumn{2}{c}{$k_BT=0.9\varepsilon$}\\
& Adv & Rec &  Adv & Rec \\\midrule
40 grid points per direction, $L_1 = 4\sigma,6\sigma$ for $k_BT=0.75\varepsilon$ and $0.9\varepsilon$, respectively, $L_2 = 2\sigma$ (as in Fig.~\ref{fig:Nano:DDFT:CheckMKT})
& 9.4653 & 4.5534 & 2.3737 & 1.0345 \\\hline
60 grid points per direction,
$L_1 = 6\sigma,8\sigma$ for $k_BT=0.75\varepsilon$ and $0.9\varepsilon$, respectively, $L_2 =3\sigma$ (figures at the top)
 &  9.7855 & 4.5888 &2.3692 & 1.0390  \\\hline
80 grid points per direction,
$L_1 = 8\sigma,10\sigma$ for $k_BT=0.75\varepsilon$ and $0.9\varepsilon$, respectively, $L_2 =4\sigma$
 &  9.8246 & 4.6023 &2.3675 & 1.0396  \\\bottomrule
\end{tabular}
\caption{Robustness of results to domain size and resolution. Top, figures: Parameters and symbols as in
Fig.~\ref{fig:Nano:DDFT:CheckMKT}, but with a finer numerical grid
with 60 points in each direction. The black solid lines are fits
to the prediction of Eq.~(\ref{eq:Nano:DDFT:MKTmodel}). Bottom, table: The friction coefficients
$\frictionCoefficientMKT$ obtained as fitting parameters of the black lines
in the figure are compared with the results for lower resolutions, in
multiples of $\sqrt{\varepsilon m}/\sigma^2$.
 \label{fig:FineNumericalGrid}}
\end{figure}

Independence of the results on the numerical grid and domain is shown in
Fig.~\ref{fig:FineNumericalGrid} using the contact line friction as a
benchmark (contact line friction is discussed in detail in
Sec.~\ref{emergent-meso} and detailed computations of this parameter are
given in
Figs.~\ref{fig:MKT:Robustness:adv:kbT075}-\ref{fig:MKT:Robustness:rec:kbT09}
of Appendix~\ref{contactlinefriccomp}). The initial configuration is a flat
liquid-vapor interface at a contact angle $\thInitial$ to the wall
\begin{equation*}
\nDensityInitial(y_1,y_2;\thInitial) = \nDensityWL(y_2)
\expandafter\hat\nDensityLV(z) + \nDensityWV(y_2) \klamm{1 -
\expandafter\hat\nDensityLV(z)},
\end{equation*}
where $\nDensityWL(y_2)$ and $\nDensityWV(y_2)$ are the equilibrium density
distributions of wall-liquid and wall-vapor interfaces, respectively,
$\expandafter\hat\nDensityLV \defi \klamm{\nDensityLV(z) -
\nDensityV}/\klamm{\nDensityL - \nDensityV}$ with $z
\defi y_2 \cos \thInitial - y_1 \sin \thInitial$.
As is evident from the figure, the scheme converges quickly with both the
number of grids and computational domain.

\subsection{Extracting contact angle, velocity and contact line friction}

In Fig.~\ref{fig:Nano:DDFT:CheckMKT}, we plot the contact line velocity as a
function of the force acting on the contact angle, effectively the difference
between the contact angle and its equilibrium value, and extract pertinent
information about the contact line friction. Monitoring the evolution of the
contact line and contact angle in Figs.~\ref{fig:Overview:kBT075} and
\ref{fig:Overview:kBT09} of Appendix~\ref{sec:dyn-evol-FlowPro} ensures that
the simulations are in a regime where any initial transients have died out
and where temporal oscillations are reduced to an acceptable level.
Typically, any transients fade away relatively quickly, at times of the order
of a few of tens of $\tau$ at the most. Additionally, the near-wall density
oscillations must be taken into account and an effective interface must be
calculated as described in the caption of Fig.~\ref{fig:Overview:kBT075} of
Appendix~\ref{sec:dyn-evol-FlowPro}.

The position of the contact line is determined by (dynamically) fitting the
isoline for the density $(\nDensity - \nDensityV)/(\nDensityL - \nDensityV) =
0.5$ to the line $(\CLPos + \frac{y_2}{\tan \theta},y_2)$ in an interval
$\mathcal{I}$ from the wall, with the contact line position $y_{1,0}$ and the
inner region contact angle $\theta$ as fitting parameters. Once the contact
line position, $y_{1,0}$, is determined, contact line velocity, $\CLVel$, is
defined as $\CLVel \defi \frac{\dI}{\dI t} \CLPos$. In
Figs.~\ref{fig:MKT:Robustness:adv:kbT075}-\ref{fig:MKT:Robustness:rec:kbT09}
of Appendix~\ref{contactlinefriccomp} we study the sensitivity of the contact
angle and contact line velocity, $\CLVel$, on the choice of the interval
$\mathcal{I}$ for the four cases depicted in
Fig.~\ref{fig:shearCompressionReceding}.

An interval that is too close to the contact line is impacted by the highly
oscillatory structure of the density profile there, and therefore leads to
higher variability. In contrast, choosing an interval which is too wide, or
too far away from the contact line, fails to capture the nanoscale dynamics.
As discussed in Sec.~\ref{emergent-meso}, plotting the contact line velocity
$U_{CL}$ versus the force acting on the contact line, $\gamma \klamm{\cos
\theta - \cos \thYoung}$, reveals an approximately linear dependency. If not
stated otherwise, we employed intervals $\mathcal{I} = [4.5\sigma,7.5\sigma]$
and $[8.5\sigma,11.5\sigma]$ for temperatures $k_BT = 0.75\varepsilon$ and
$0.9\varepsilon$, respectively.

\subsection{Computational cost}

The computations were done on a Windows Intel(R) Core(TM) i7-6700 CPU @3.40
GHz with 16.0GB RAM using MATLAB R2016a and take approximately 100 mins to
compute the convolution matrices for one contact angle setting, 20 mins for
one equilibrium and 10-15 mins for one dynamic computation.

\section{Dynamic evolution of receding and advancing contact lines and flow properties\label{sec:dyn-evol-FlowPro}}
\begin{figure}
\centering		
\footnotesize
\begin{tabular}{c|ccccc}
\rotatebox[origin=c]{90}{$\theta_{\text{in}} \to \theta_{\text{eq}}$}
& $t = 0\tau$ & $t = 250\tau$ & Liquid-vapor interface & Contact angle & Contact line velocity \\ \midrule

\rotatebox[origin=c]{90}{$90^\circ \to 60^\circ$} &
\raisebox{-.5\height}{\hspace{-0.1cm}
  \includegraphics[height=2.8cm]{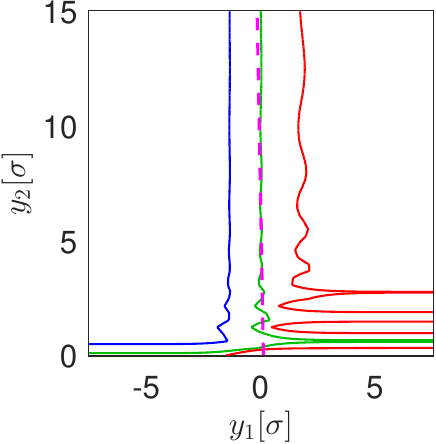}} &
\raisebox{-.5\height}{\hspace{-0.1cm}
  \includegraphics[height=2.8cm]{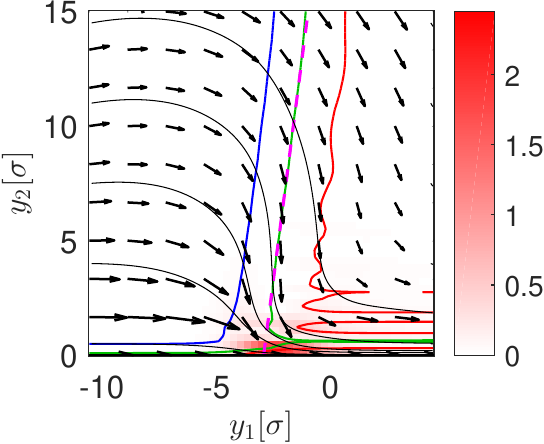}} &
\raisebox{-.5\height}{\hspace{-0.1cm}
  \includegraphics[height=2.8cm]{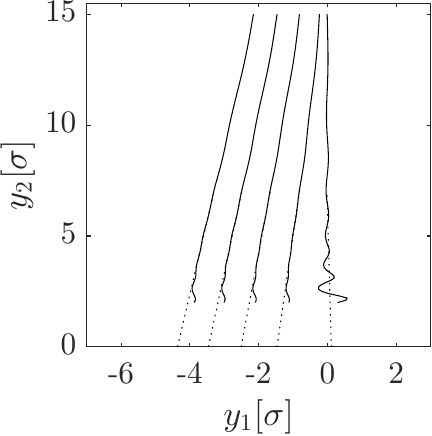}} &
\raisebox{-.5\height}{\hspace{-0.1cm}
  \includegraphics[height=2.8cm]{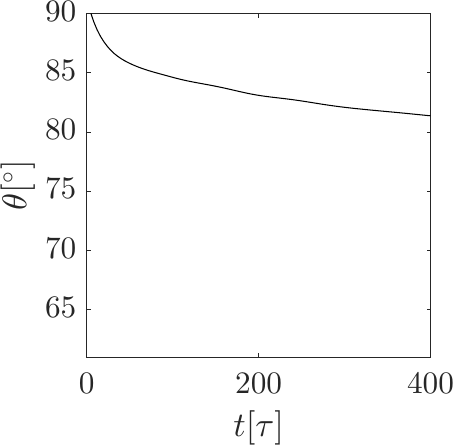}} &
\raisebox{-.5\height}{\hspace{-0.1cm}
  \includegraphics[height=2.8cm]{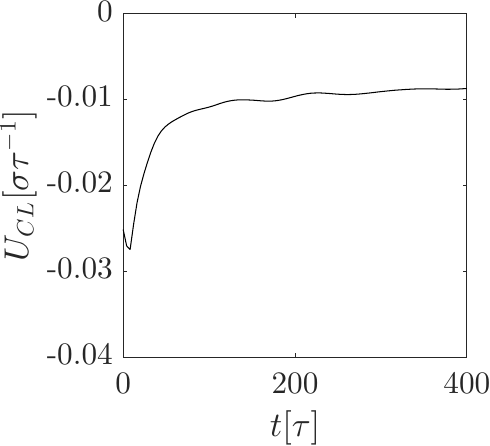}} \\\addlinespace[4pt]

\rotatebox[origin=c]{90}{$90^\circ \to 120^\circ$} &
\raisebox{-.5\height}{\hspace{-0.1cm}
  \includegraphics[height=2.8cm]{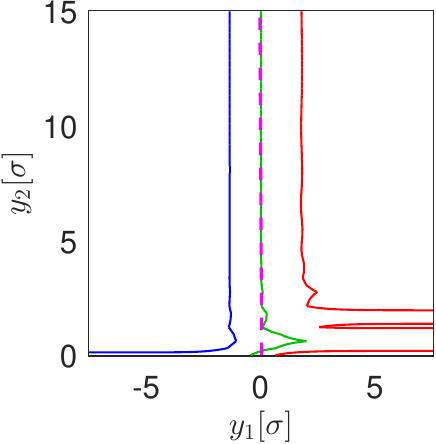}} &
\raisebox{-.5\height}{\hspace{-0.1cm}
  \includegraphics[height=2.8cm]{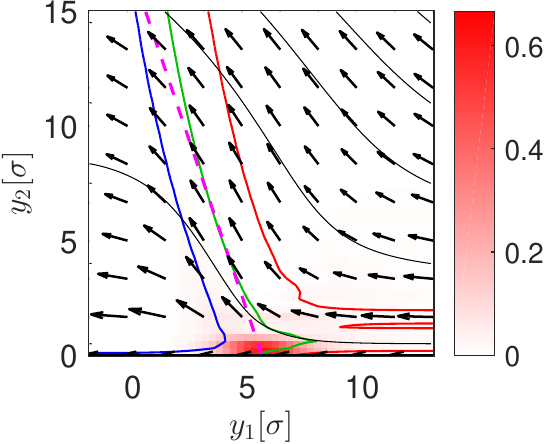}} &
\raisebox{-.5\height}{\hspace{-0.1cm}
  \includegraphics[height=2.8cm]{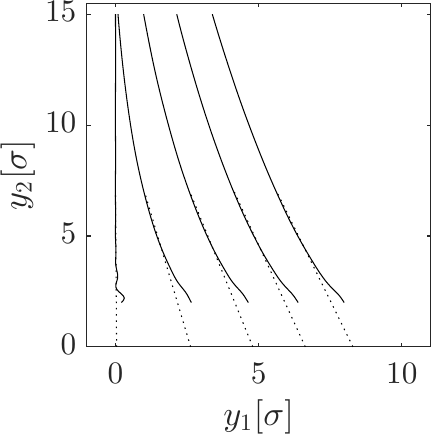}} &
\raisebox{-.5\height}{\hspace{-0.1cm}
  \includegraphics[height=2.8cm]{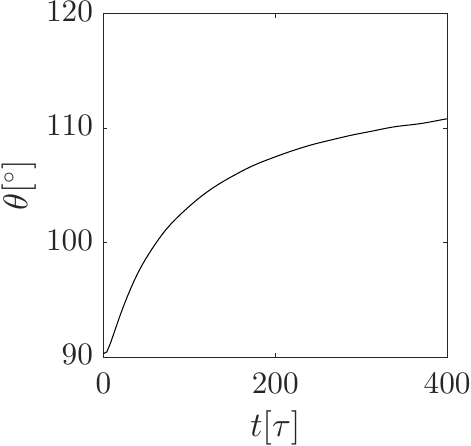}} &
\raisebox{-.5\height}{\hspace{-0.1cm}
  \includegraphics[height=2.8cm]{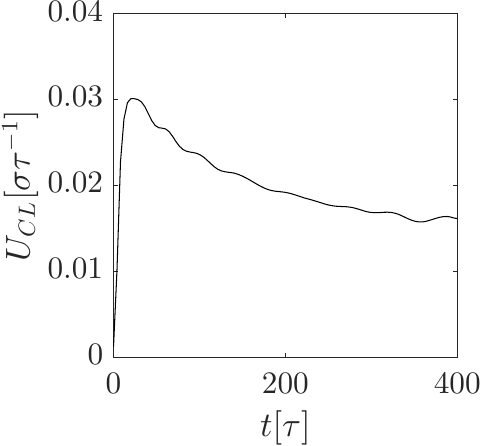}} \\\addlinespace[4pt]

\rotatebox[origin=c]{90}{$120^\circ \to 90^\circ$} &
\raisebox{-.5\height}{\hspace{-0.1cm}
  \includegraphics[height=2.8cm]{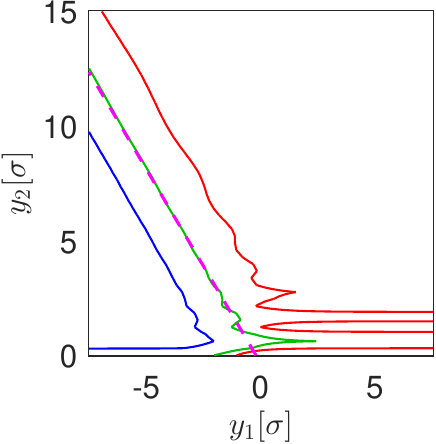}} &
\raisebox{-.5\height}{\hspace{-0.1cm}
  \includegraphics[height=2.8cm]{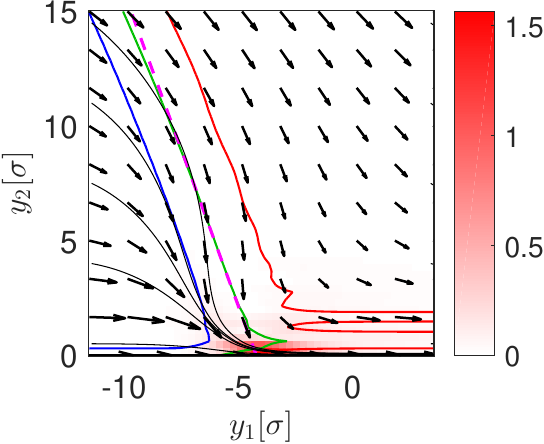}} &
\raisebox{-.5\height}{\hspace{-0.1cm}
  \includegraphics[height=2.8cm]{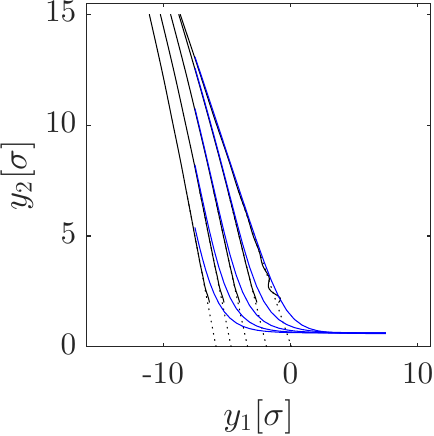}} &
\raisebox{-.5\height}{\hspace{-0.1cm}
  \includegraphics[height=2.8cm]{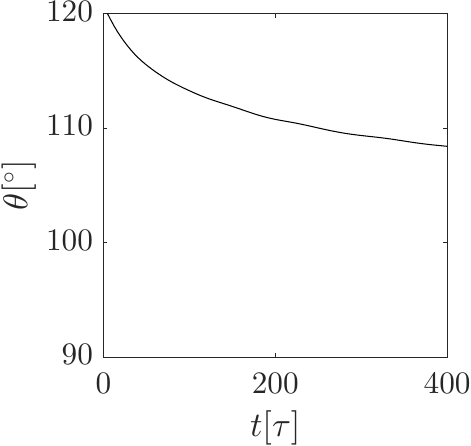}} &
\raisebox{-.5\height}{\hspace{-0.1cm}
  \includegraphics[height=2.8cm]{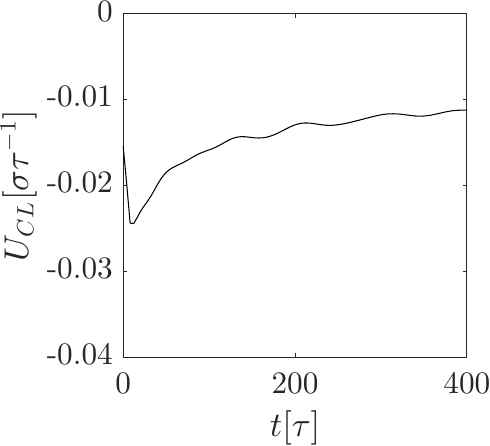}} \\\addlinespace[4pt]

\rotatebox[origin=c]{90}{$60^\circ \to 90^\circ$} &
\raisebox{-.5\height}{\hspace{-0.1cm}
  \includegraphics[height=2.8cm]{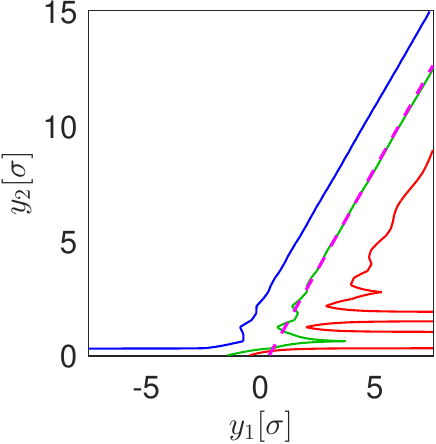}} &
\raisebox{-.5\height}{\hspace{-0.1cm}
  \includegraphics[height=2.8cm]{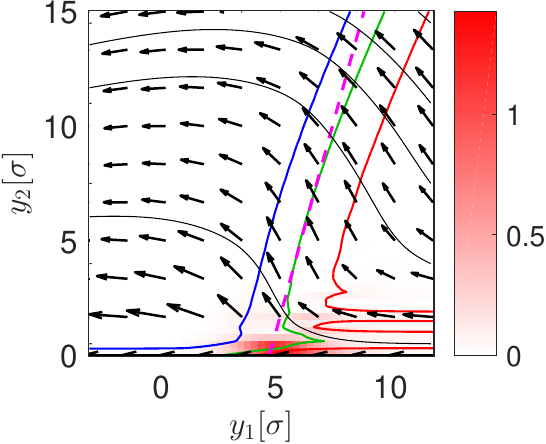}} &
\raisebox{-.5\height}{\hspace{-0.1cm}
  \includegraphics[height=2.8cm]{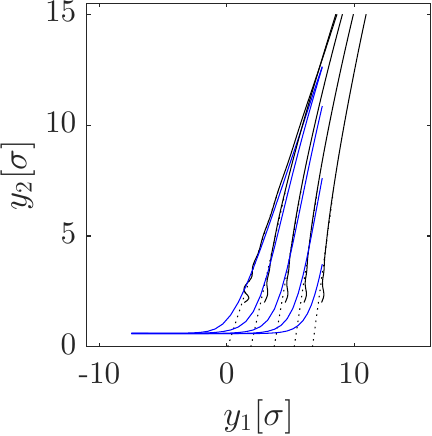}} &
\raisebox{-.5\height}{\hspace{-0.1cm}\includegraphics[height=2.8cm]{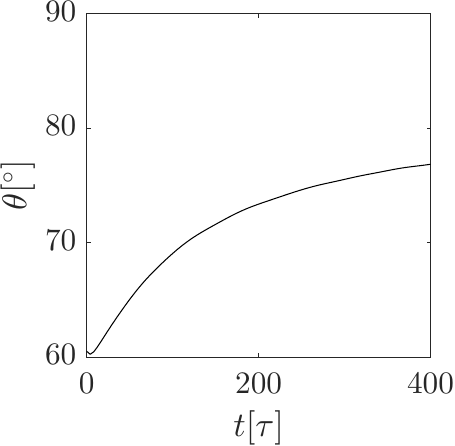}} &
\raisebox{-.5\height}{\hspace{-0.1cm}\includegraphics[height=2.8cm]{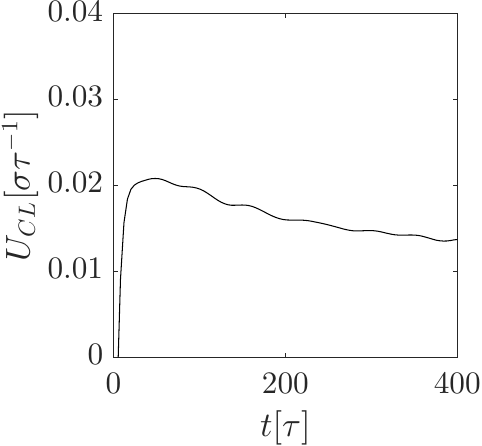}} \\\addlinespace[4pt]

\rotatebox[origin=c]{90}{$60^\circ \to 0^\circ$} &
\raisebox{-.5\height}{\hspace{-0.1cm}
  \includegraphics[height=2.8cm]{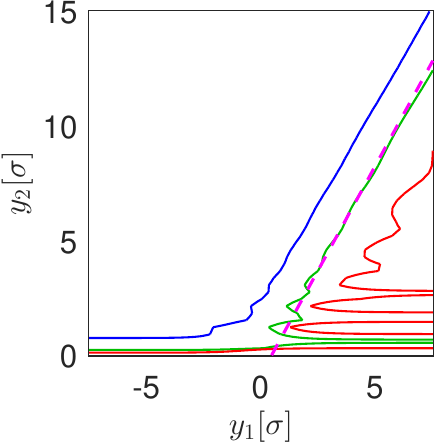}} &
\raisebox{-.5\height}{\hspace{-0.1cm}
  \includegraphics[height=2.8cm]{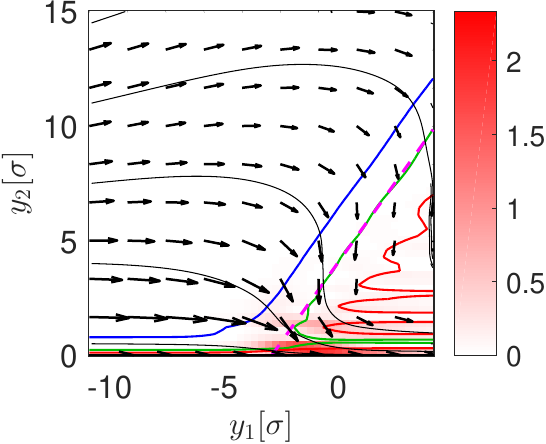}} &
\raisebox{-.5\height}{\hspace{-0.1cm}
  \includegraphics[height=2.8cm]{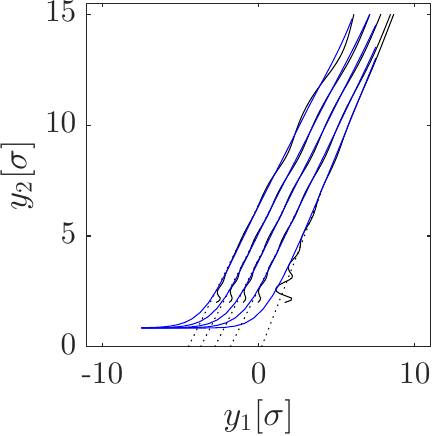}} &
\raisebox{-.5\height}{\hspace{-0.1cm}
  \includegraphics[height=2.8cm]{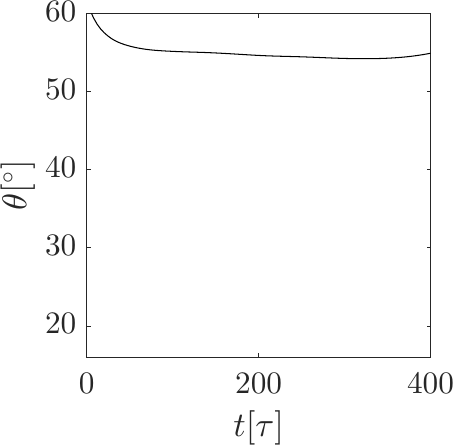}} &
\raisebox{-.5\height}{\hspace{-0.1cm}
  \includegraphics[height=2.8cm]{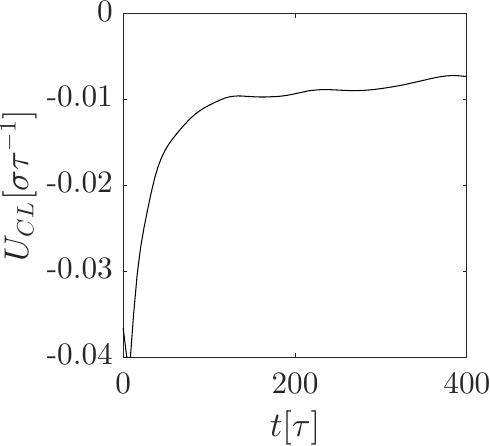}} \\

\end{tabular}
\caption{Contact line behavior for different wetting and initial conditions for $k_B T = 0.75\varepsilon$.
Columns I and II: Density isolines, streamlines, and viscous energy dissipation at the moving contact line at the start of the simulation and at $t = 250\tau$. $y_{1,2}$ are the streamwise and cross-stream coordinates, respectively. Column III: Position of the liquid-vapor interface defined as the position where $n = (\nDensityL + \nDensityV)/2$, at time points $t = 0,100,200,300,400\tau$ (black lines). Dotted lines are the linear interpolation of the liquid-vapor interface. For initial contact angles $\theta \neq 90^\circ$, the adsorption film thickness is represented by blue lines. Columns IV and V: Contact angle and contact line velocity over time, as obtained from the extrapolation of the dotted lines in column III. Velocities, velocity fields, and streamlines are given relative to the motion of the contact line. \label{fig:Overview:kBT075}}
\end{figure}

\begin{figure}
\centering		
\footnotesize
\begin{tabular}{c|ccccc}
\rotatebox[origin=c]{90}{$\theta_{\text{in}} \to \theta_{\text{eq}}$}
& $t = 0\tau$ & $t = 250\tau$ & Liquid-vapor interface & Contact angle & Contact line velocity \\ \midrule

\rotatebox[origin=c]{90}{$90^\circ \to 60^\circ$} &
\raisebox{-.5\height}{\hspace{-0.1cm}
  \includegraphics[height=2.8cm]{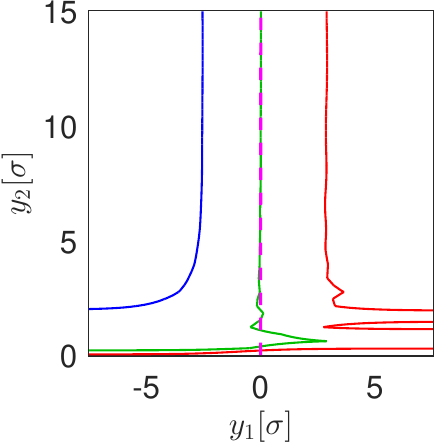}} &
\raisebox{-.5\height}{\hspace{-0.1cm}
  \includegraphics[height=2.8cm]{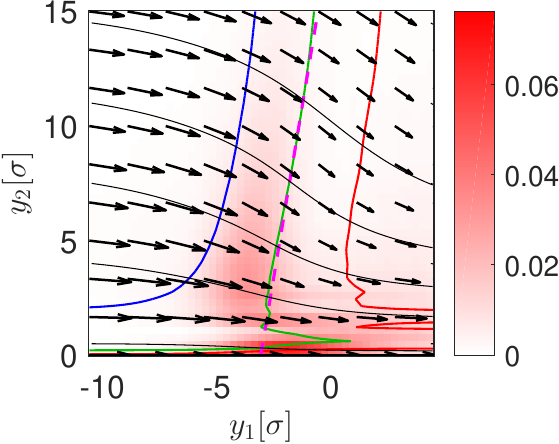}} &
\raisebox{-.5\height}{\hspace{-0.1cm}
  \includegraphics[height=2.8cm]{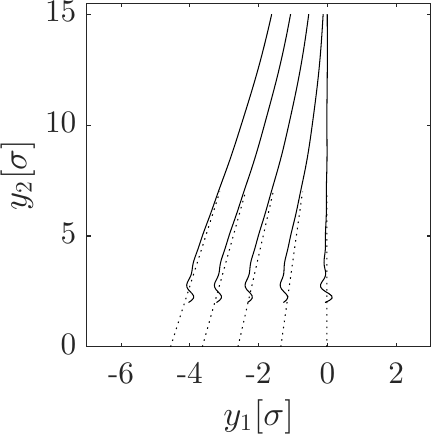}} &
\raisebox{-.5\height}{\hspace{-0.1cm}
  \includegraphics[height=2.8cm]{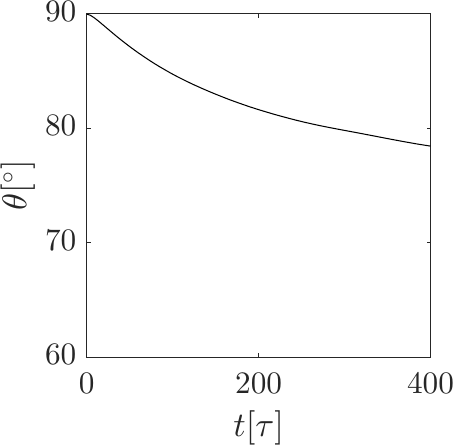}} &
\raisebox{-.5\height}{\hspace{-0.1cm}
  \includegraphics[height=2.8cm]{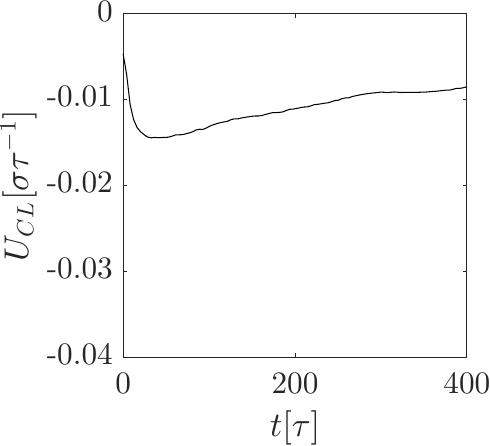}} \\\addlinespace[4pt]

 \rotatebox[origin=c]{90}{$90^\circ \to 120^\circ$} &
 \raisebox{-.5\height}{\hspace{-0.1cm}\includegraphics[height=2.8cm]{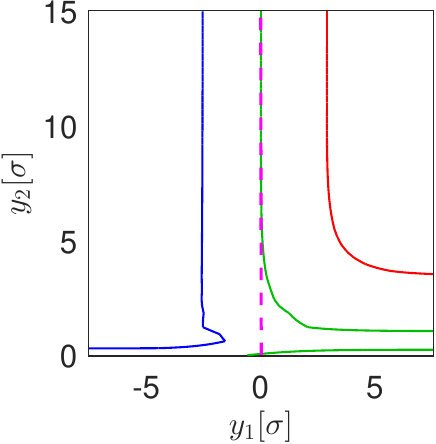}} &
 \raisebox{-.5\height}{\hspace{-0.1cm}\includegraphics[height=2.8cm]{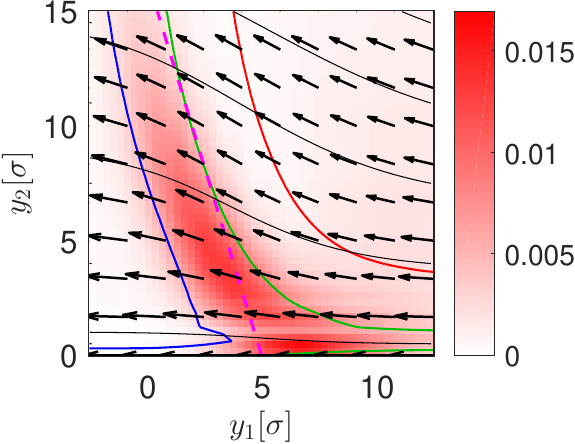}} &
 \raisebox{-.5\height}{\hspace{-0.1cm}\includegraphics[height=2.8cm]{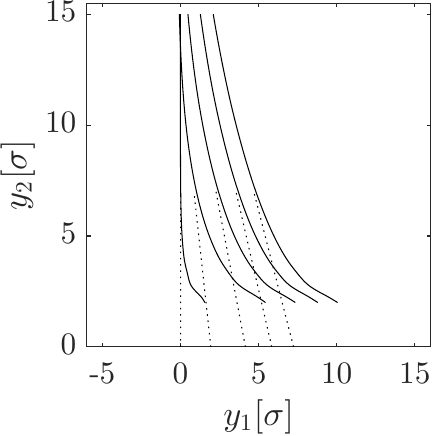}} &
 \raisebox{-.5\height}{\hspace{-0.1cm}\includegraphics[height=2.8cm]{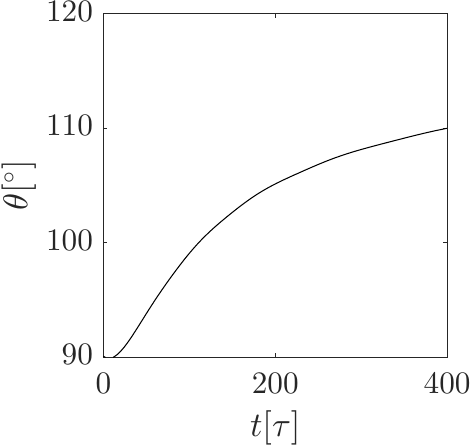}} &
 \raisebox{-.5\height}{\hspace{-0.1cm}\includegraphics[height=2.8cm]{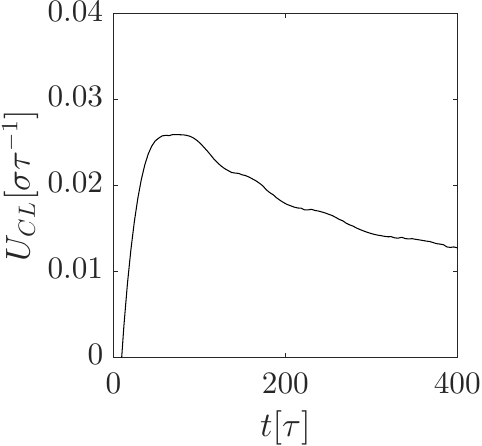}} \\\addlinespace[4pt]

 \rotatebox[origin=c]{90}{$120^\circ \to 90^\circ$} &
 \raisebox{-.5\height}{\hspace{-0.1cm}\includegraphics[height=2.8cm]{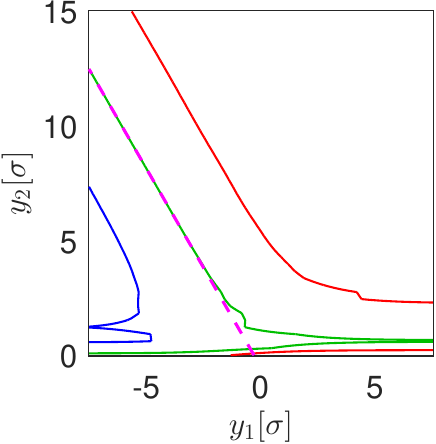}} &
 \raisebox{-.5\height}{\hspace{-0.1cm}\includegraphics[height=2.8cm]{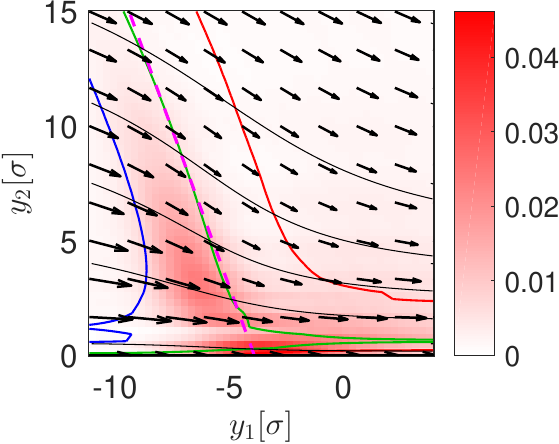}} &
 \raisebox{-.5\height}{\hspace{-0.1cm}\includegraphics[height=2.8cm]{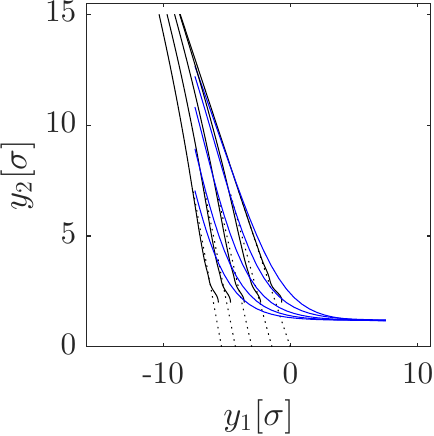}} &
 \raisebox{-.5\height}{\hspace{-0.1cm}\includegraphics[height=2.8cm]{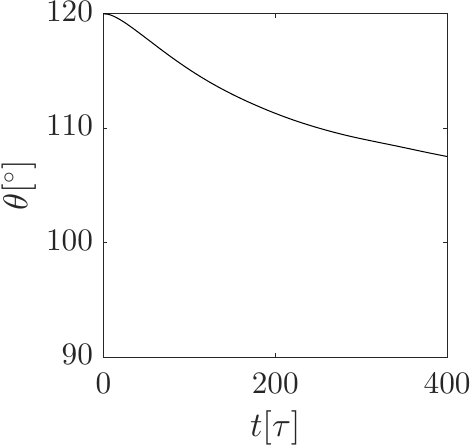}} &
 \raisebox{-.5\height}{\hspace{-0.1cm}\includegraphics[height=2.8cm]{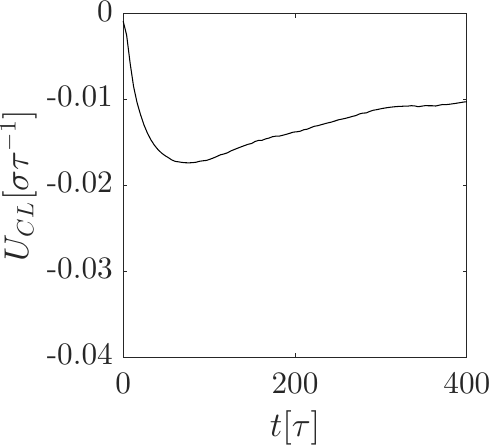}} \\\addlinespace[4pt]

 \rotatebox[origin=c]{90}{$60^\circ \to 90^\circ$} &
 \raisebox{-.5\height}{\hspace{-0.1cm}\includegraphics[height=2.8cm]{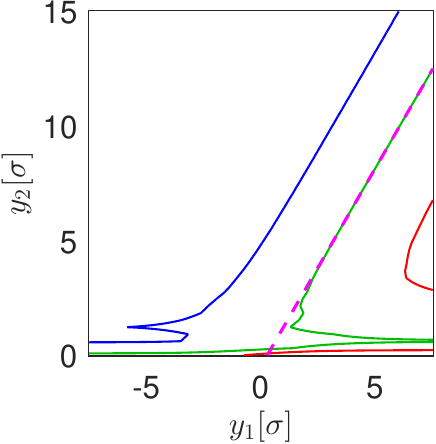}} &
 \raisebox{-.5\height}{\hspace{-0.1cm}\includegraphics[height=2.8cm]{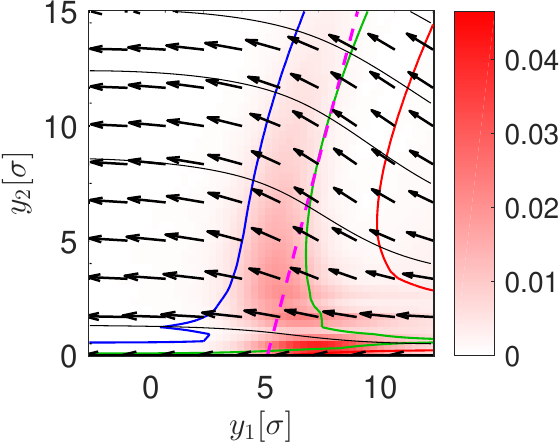}} &
 \raisebox{-.5\height}{\hspace{-0.1cm}\includegraphics[height=2.8cm]{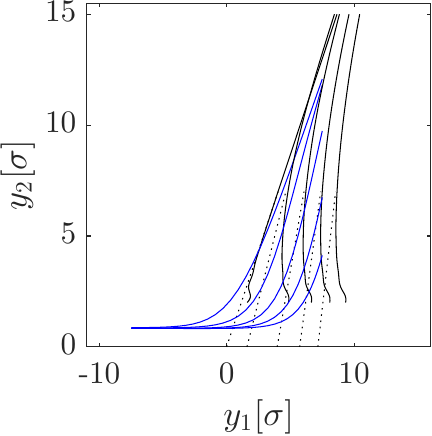}} &
 \raisebox{-.5\height}{\hspace{-0.1cm}\includegraphics[height=2.8cm]{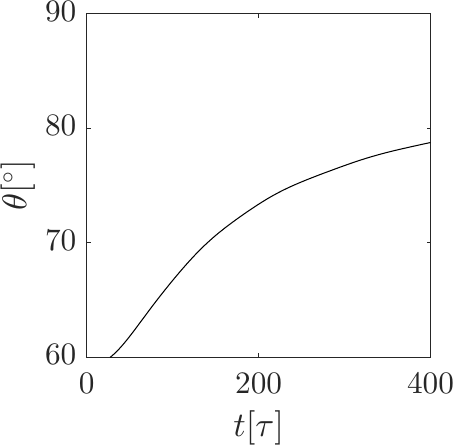}} &
 \raisebox{-.5\height}{\hspace{-0.1cm}\includegraphics[height=2.8cm]{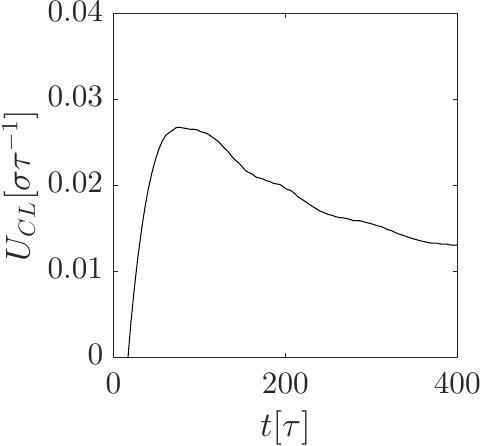}} \\\addlinespace[4pt]

 \rotatebox[origin=c]{90}{$60^\circ \to 0^\circ$} &
 \raisebox{-.5\height}{\hspace{-0.1cm}\includegraphics[height=2.8cm]{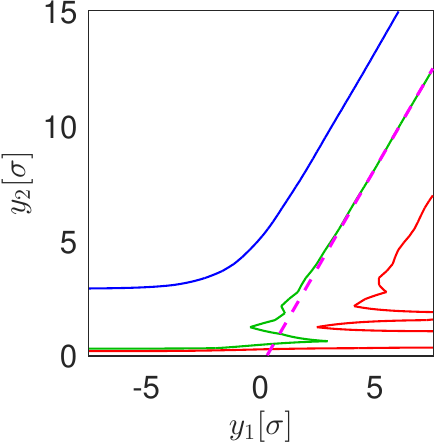}} &
 \raisebox{-.5\height}{\hspace{-0.1cm}\includegraphics[height=2.8cm]{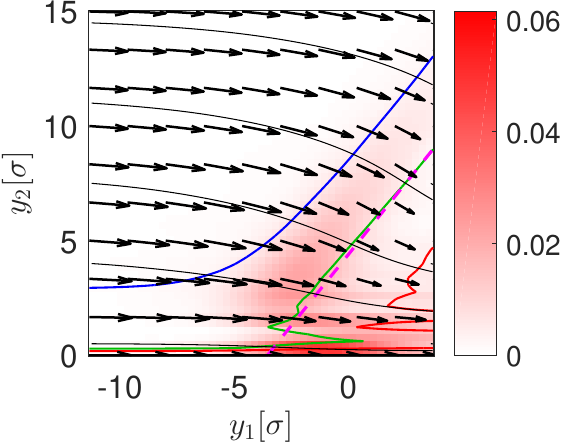}} &
 \raisebox{-.5\height}{\hspace{-0.1cm}\includegraphics[height=2.8cm]{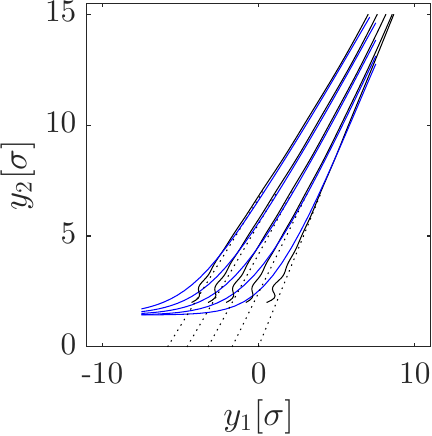}} &
 \raisebox{-.5\height}{\hspace{-0.1cm}\includegraphics[height=2.8cm]{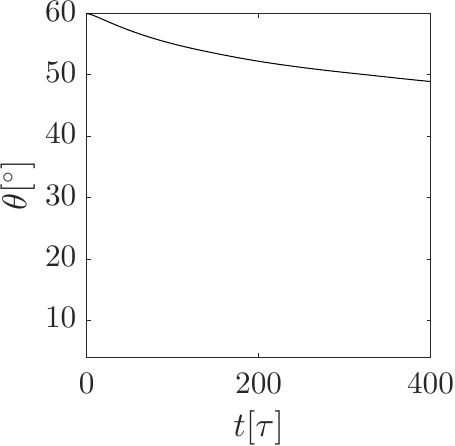}} &
 \raisebox{-.5\height}{\hspace{-0.1cm}\includegraphics[height=2.8cm]{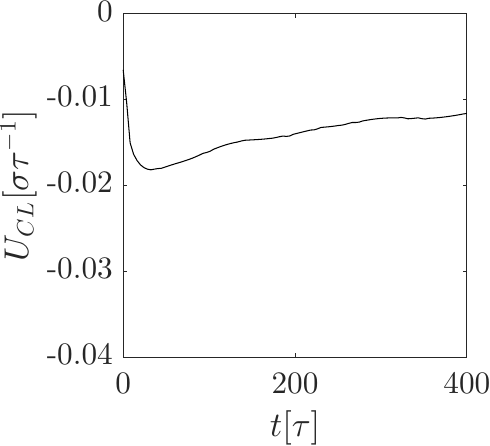}} \\

\end{tabular}
\caption{Data as in Fig.~\ref{fig:Overview:kBT075}, but for $k_B T = 0.9\varepsilon$. \label{fig:Overview:kBT09}}
\end{figure}

Figures~\ref{fig:Overview:kBT075} and~\ref{fig:Overview:kBT09} provide a
detailed overview of contact line behavior for different wetting and initial
conditions for two different temperatures, $k_B T = 0.75\varepsilon$ and $k_B
T = 0.9\varepsilon$, respectively.

\newcommand{\figHeightFirst}{2.8cm}

\newcommand{\IncludeSnapshotStart}[2]{
 \raisebox{-.5\height}{\hspace{-0.1cm} \includegraphics[height=\figHeightFirst]{Figures/SI_Fig1/#1/Dynamics/#2__entropy_rho_t_fittedInterface_contactangle_0_t_0}}}

\newcommand{\IncludeSnapshotEnd}[2]{
 \raisebox{-.5\height}{\hspace{-0.1cm} \includegraphics[height=\figHeightFirst]{Figures/SI_Fig1/#1/Dynamics/#2__entropy_rho_t_fittedInterface_UV_t_contactangle_0_t_250}}}

 \newcommand{\IncludeContactAngle}[2]{\raisebox{-.5\height}{\hspace{-0.1cm} \includegraphics[height=\figHeightFirst]{Figures/SI_Fig1/#1/Dynamics/#2_contactAngle}}}
 \newcommand{\IncludeContactLineVelocity}[2]{\raisebox{-.5\height}{\hspace{-0.1cm} \includegraphics[height=\figHeightFirst]{Figures/SI_Fig1/#1/Dynamics/#2_CLVelocity}}}
 \newcommand{\IncludeInterfaceFitting}[2]{\raisebox{-.5\height}{\hspace{-0.1cm} \includegraphics[height=\figHeightFirst]{Figures/SI_Fig1/#1/Dynamics/#2_InterfaceFitting}}}

\newcommand{\IncludeRow}[2]{
 \IncludeSnapshotStart{#1}{#2} &
 \IncludeSnapshotEnd{#1}{#2}  &
  \IncludeInterfaceFitting{#1}{#2}  &
 \IncludeContactAngle{#1}{#2}  &
 \IncludeContactLineVelocity{#1}{#2}
}



\begin{figure}[h!tp]
\centering
\begin{tabular}{cc}
Advancing & Receding\\
\includegraphics[width=0.47\linewidth]{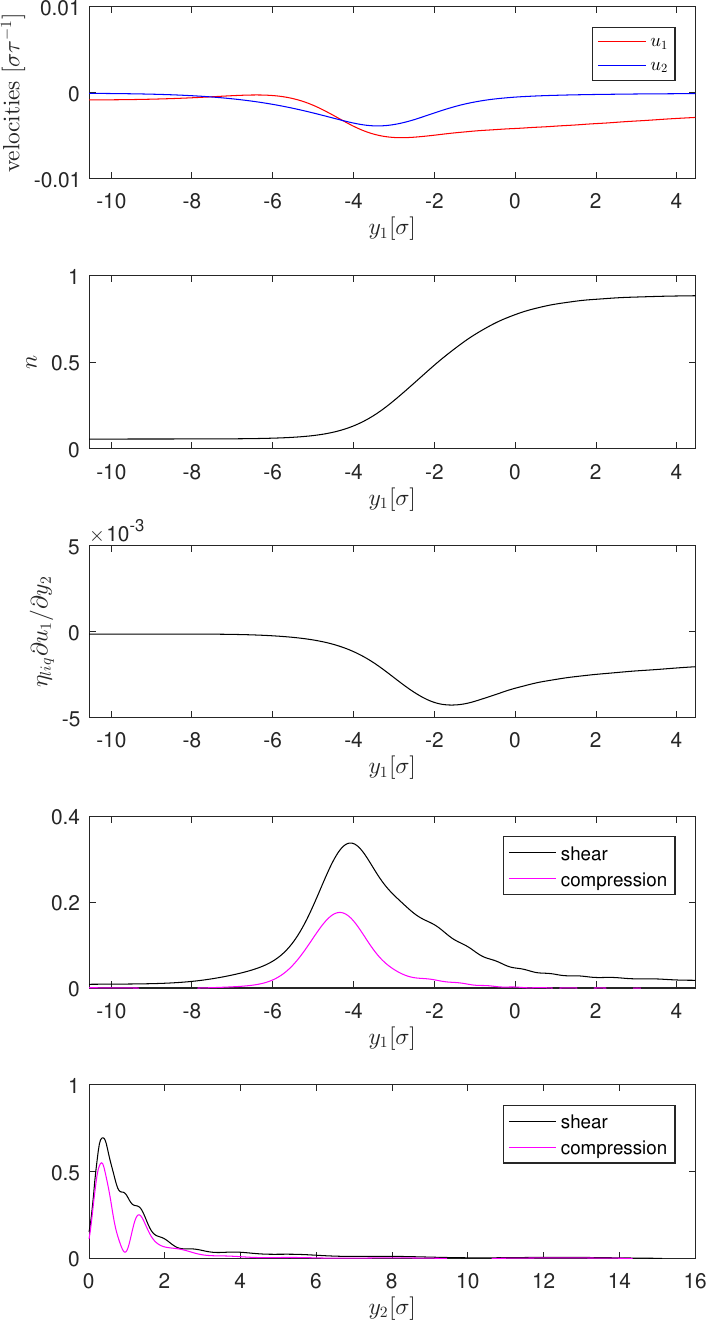} &
\includegraphics[width=0.47\linewidth]{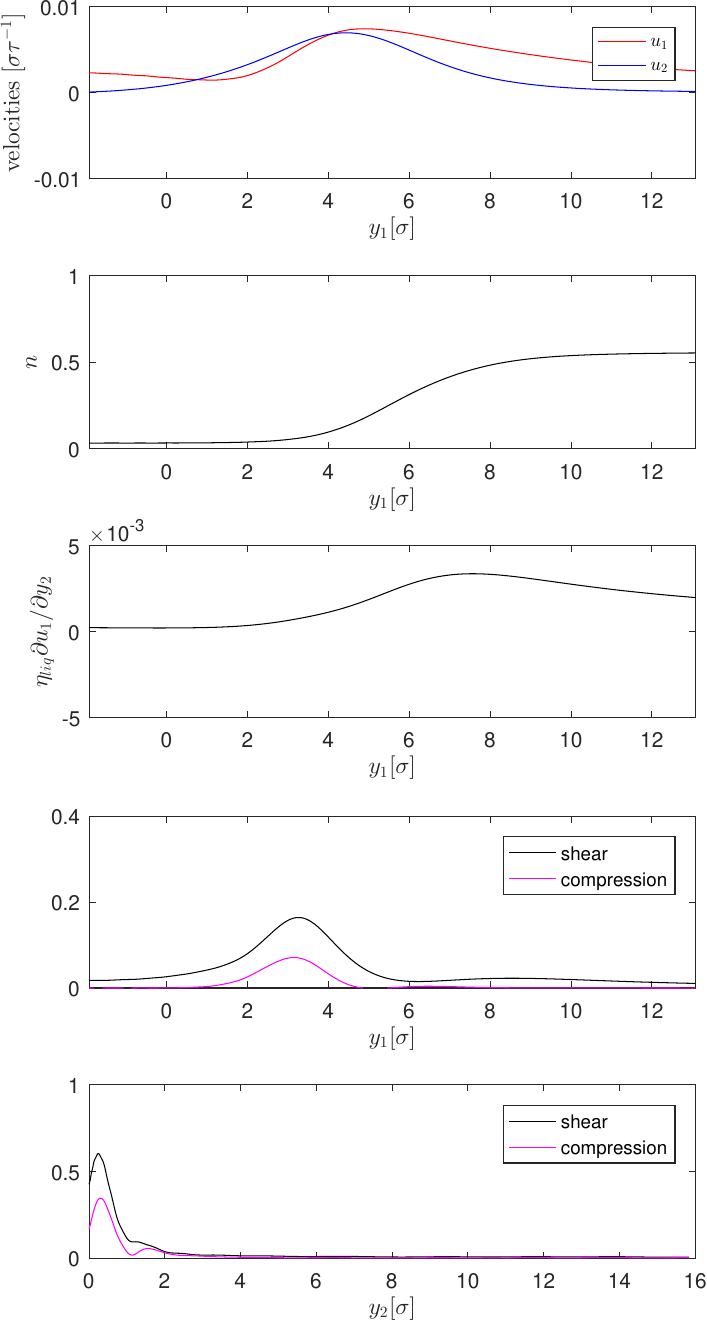}
\end{tabular}
\caption{Flow properties in the vicinity of the contact line for advancing and receding contact lines at $k_BT= 0.75 \varepsilon$ and at $t=250\tau$,
with parameters as in Fig.~\ref{fig:shearCompressionReceding}. The first four rows depict velocities $u_{1,2}$ in the $y_{1,2}$ directions,
respectively, number density $n$, shear force, and shear and compressive effects parallel to the wall, at a distance of $1.1\sigma$ from the wall,
corresponding to the bracketed quantities in Eq.~\ref{eq:ViscousHeatProduction}.
The bottom row depicts shear and compressive contributions parallel to the liquid-vapor interface, at a distance of $1\sigma$ into the vapor phase,
again corresponding to the bracketed quantities in Eq.~\ref{eq:ViscousHeatProduction}.
The bottom two rows are analogous to Fig.~\ref{fig:shearCompressionReceding:2}. \label{fig:FlowPropertiesAroundTheCL}}
\end{figure}

\begin{figure}[h!tp]
\centering
\begin{tabular}{cc}
Advancing & Receding\\
\includegraphics[width=0.47\linewidth]{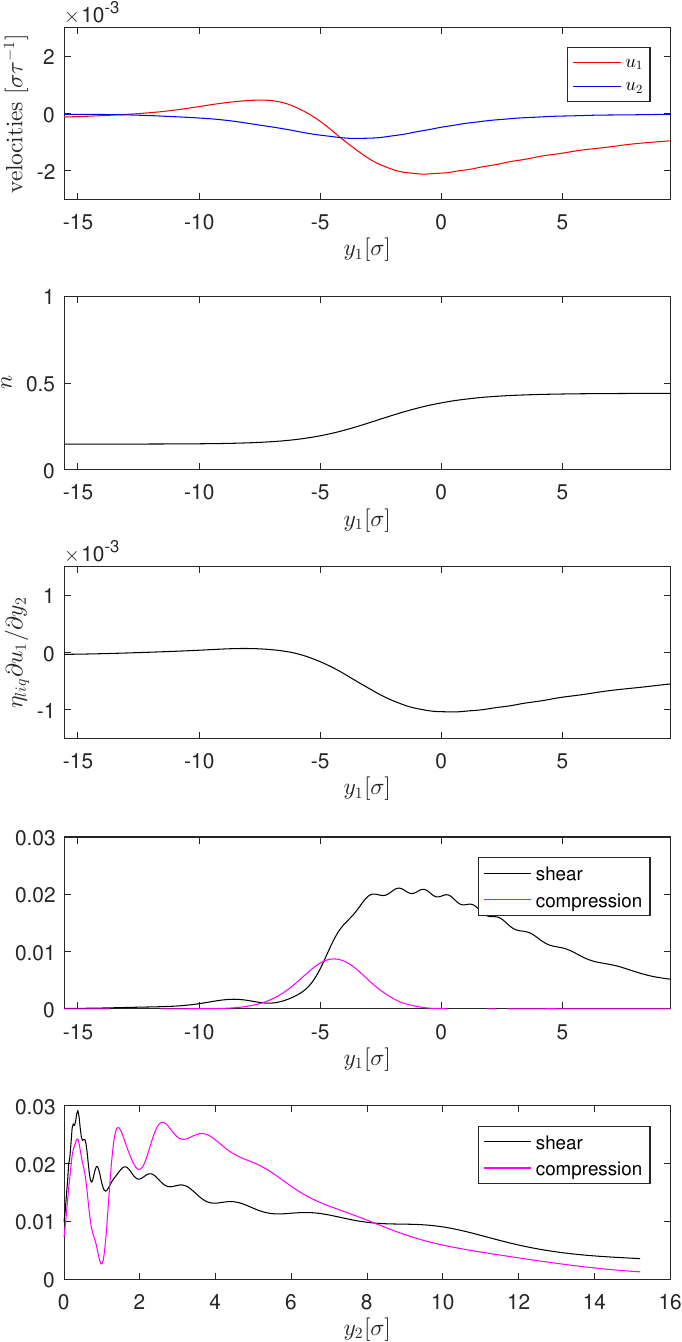} &
\includegraphics[width=0.47\linewidth]{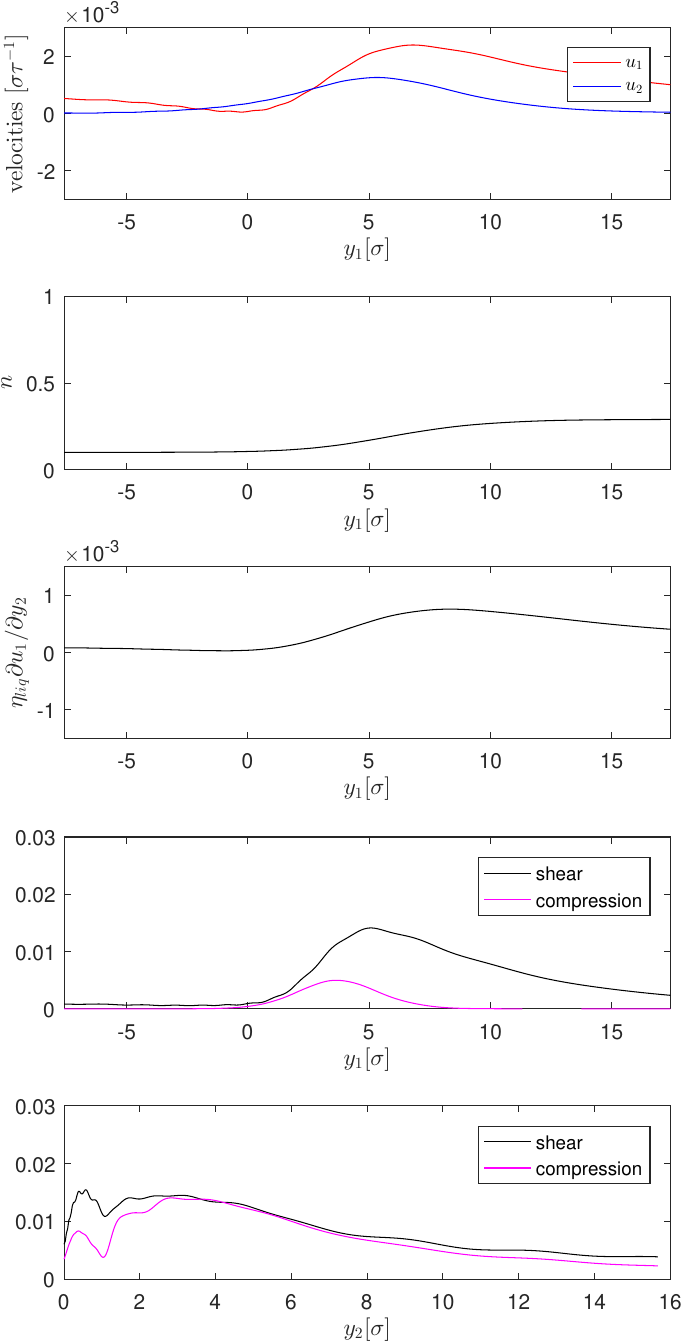}
\end{tabular}
\caption{Flow properties in the vicinity of the contact line for advancing and receding contact lines at $k_BT= 0.9 \varepsilon$ and at $t=250\tau$, analogous to
Fig.~\ref{fig:FlowPropertiesAroundTheCL}. \label{fig:FlowPropertiesAroundTheCL:09}}
\end{figure}

Figures~\ref{fig:FlowPropertiesAroundTheCL}
and~\ref{fig:FlowPropertiesAroundTheCL:09} display in detail a range of flow
properties in the vicinity of the contact line for both advancing and
receding contact lines.

\newcommand{\SizeSlipFigures}{1.0in}

\section{Effect of changing viscosity on contact line motion\label{viscochange}}

Here we explore the dependency of the contact line motion on shear and
compressive effects shown in Figs~\ref{fig:Fig3_SimulationsMovingCA}
and~\ref{fig:shearCompressionReceding}. For this purpose we varied the shear
and bulk viscosities, $\shearViscosity$ and $\bulkViscosity$, respectively.
Changes in contact line speed, shown in Tab.~\ref{tab:ViscosityVariation},
inform us on the dependence of contact line motion on each effect. This leads
to the following general observations:
\begin{enumerate}
\item Contact line motion is more sensitive to changes of liquid than vapor
    shear viscosity $\shearViscosity$. On the contrary, the sensitivity for
    bulk viscosity $\bulkViscosity$ is reversed, and contact line motion is
    more sensitive to changes in the vapor rather than in the liquid side.
    This result is consistent throughout all computations. It is in line
    with the location of the effective shear layer in the liquid side of
    the contact line and the peak of the compression in the vapor side of
    the contact line (see Fig.~\ref{fig:shearCompressionReceding:2}).
\item In general, advancing contact lines are more sensitive to changes in
    viscosity. This is especially pronounced for changes of the bulk
    viscosity in the low temperature regime ($k_BT = 0.75 \varepsilon$).
\item In general, the low temperature regime is more sensitive to changes
    in viscosity compared to the high temperature regime. This is in line
    with the localization of the energy dissipation at the vicinity of the
    contact line for low temperatures. However, this trend is non-existent
    or reversed as far as the effect of compressive viscosity on receding
    contact lines is concerned.
\item Changes in the liquid and vapor side can be superimposed to get the
    total change, i.e. increasing one viscosity first and then the other,
    is equivalent to increasing both at the same time.
\item The sensitivity of contact line motion to increasing vs decreasing of
    the liquid side viscosities is the same.
\end{enumerate}

\begin{table}[htp]
\resizebox{\textwidth}{!}{%
\centering
\begin{tabular}{ccc|c||cccc|cccc}
&&&& \multicolumn{4}{c}{Shear viscosity}  &  \multicolumn{4}{c}{Bulk viscosity}  \\
$k_BT/\varepsilon$ & & Target CA & $\Delta t$, standard & $\shearViscosityL,\shearViscosityV \uparrow$	 & $\shearViscosityL \uparrow$	 &
$\shearViscosityV \uparrow$
	 & $\shearViscosityL \downarrow$ &	
	  $\bulkViscosityL,\bulkViscosityV \uparrow$	 & $\bulkViscosityL \uparrow$	 &
$\bulkViscosityV \uparrow$
	 & $\bulkViscosityL \downarrow$ \\\hline
	 $0.75$	& Adv&	$83^\circ$ &	$207.1$	        & $28.5\%$	& $15.3\%$	&$11.5\%$	& $-14.1\%$& $54.4\%$	& $21.3\%$	& $33.2\%$	&$-13.8\%$\\
$0.75$	& Rec&	$97^\circ$	&      $43.0$		& $22.7\%$	&$13.5\%$  & $9.2\%$	& $-13.5\%$& $14.4\%$	& $4.5\%$	&$10.6\%$	&$-5.0\%$\\
$0.9$ &Adv & $83^\circ$ & $159.2$ & $18.5\%$ & $11.3\%$& $6.7\%$ & $-11.4\%$ & $19.5\%$ &	$5.7\%$	& $14.7\%$& $-6.0\%$\\
$0.9$ & Rec& $97^\circ$ &  $88.5$  & $15.1\%$ & $8.9\%$  & $5.9\%$ & $-9.1\%$	& $16.8\%$  & $5.1\%$	& $12.5\%$&$-5.5\%$
\end{tabular}}
\vspace{0.1cm}
\caption{Changes in shear and compressive viscosity differentially affect contact line motion. {\normalfont Starting from initial contact angles $\thInitial = 90^\circ$,
advancing (`Adv', $\thYoung = 60^\circ$) and receding (`Rec', $\thYoung = 120^\circ$) contact lines at two temperatures $k_BT = 0.75,0.9 \varepsilon$ are studied. The `standard'
configuration employs parameters as in Fig.~\ref{fig:Fig3_SimulationsMovingCA}.
The fourth columns shows the time $\Delta t$ in multiples of $\sigma \sqrt{m/\varepsilon}$ expired until the contact line reaches the target contact angle defined in
the third column. The remainder of the table depicts relative changes of $\Delta t$ with respect to the `standard' configuration, if the bulk and shear viscosities are increased or
decreased by $20\%$ of the difference of the respective liquid and vapor viscosities.}\label{tab:ViscosityVariation}}
\end{table}%

\section{Extracting the slip length\label{sliplength}}

\begin{figure}[h!tp]
\footnotesize
	\centering	
	\begin{tabular}{c|cc}
& Slip velocity $u_1 [\sigma\tau^{-1}]$ & Slip length $\lambda [\sigma]$\\	 \hline
\rotatebox[origin=c]{90}{Advancing CL: $90^\circ \to 60^\circ$} &
  \raisebox{-.5\height}{\includegraphics{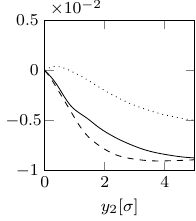}}&
  \raisebox{-.5\height}{\includegraphics{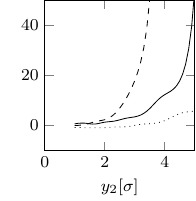}}\\
\rotatebox[origin=c]{90}{Receding CL: $90^\circ \to 120^\circ$}&
  \raisebox{-.5\height}{\includegraphics{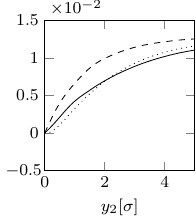}}
&  \raisebox{-.5\height}{\includegraphics{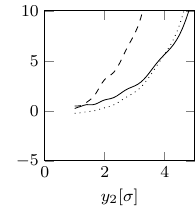}}
\end{tabular}
	\caption{Slip length computation for an advancing and receding contact line, with parameters as in the right plot of Fig.~\ref{fig:Fig3_SimulationsMovingCA}.
    Graphs in the left column depict the velocity $u_1$ in direction parallel to the wall as a function of the distance to the wall.
    The velocity is plotted along three trajectories perpendicular to the wall: crossing the contact line (solid lines),
    and shifted by $3\sigma$ towards the vapor and the liquid phase, depicted by dotted and dashed lines, respectively.
    The slip length is defined in Eq.~\ref{eq:effectiveSlipLength} and is a function of the distance to the wall.
    \label{fig:SlipVelocity}}
\end{figure}

The slip length establishes a linear relationship between the fluid velocity
and shear at the wall. Given that in our DDFT computations we impose a
no-slip condition at the wall, we can infer an effective slip length
$\lambda$ from the velocity profile at the liquid-vapor interface and at a
distance $\bar y_{2}$ to the wall:
\begin{align}
\left. \diff{u_1}{y_2}\right|_{y_2 = \bar y_{2}} \klamm{\lambda + \bar y_{2}} = {\left. u_1 \right|_{\bar y_{2}}}.
\label{eq:effectiveSlipLength}
\end{align}
$\bar y_{2}$ may be interpreted as an effective shear layer thickness.
Figure~\ref{fig:SlipVelocity} shows velocities in a direction parallel to the
wall and the slip lengths that may be extracted from those profiles as a
function of $\bar y_{2}$.

Physically, we expect that the slip length depends on the fluid density in
the immediate vicinity of the contact line which is highly oscillatory in
nature. Not surprisingly, the slip length is highly sensitive to the choice
of the position $\bar y_2$ along the wall at which the velocity $u_1$ in
Eq.~\ref{eq:effectiveSlipLength} is taken, Indeed, in
Fig.~\ref{fig:SlipVelocity} it is shown that employing the velocity $u_1$ at
three particle diameters away from the contact line into the vapor or liquid
phase along the wall significantly affects the results for the slip length.

Similar challenges arise for MD or experimental slip measurements
\cite{Ralston:2008fk}. So, whilst the initial layer of fluid particles may
well act as a slip-facilitating layer, the measurement of the slip length at
the nanoscale and subsequent insertion into macroscopic models is problematic
and does not lead to a robust prediction of macroscopic behavior. This means
that whilst slip models provide a convenient translational tool connecting
different macro- and mesoscopic velocity and contact angle measurements
\cite{Sibley:2015:CrackingAnOldNut}, their interpretation of (effective) slip
length is more attuned to that of an abstract coarse-grained parameter, as
opposed to actual physical slip at the nanoscale (see also discussion in
\cite{benilov2015thin}).

\section{Contact line friction computations\label{contactlinefriccomp}}

\begin{figure}[h!tp]
\centering
\begin{tabular}{c|ccc}
 & $w =2\sigma$ (-) & $w = 4\sigma$ (--\,--) & $w = 6\sigma$ ($\cdot\cdot\cdot$) \\ \hline
\rotatebox[origin=c]{90}{$h = 4\sigma$} &
\raisebox{-.5\height}{\includegraphics[height=3cm]{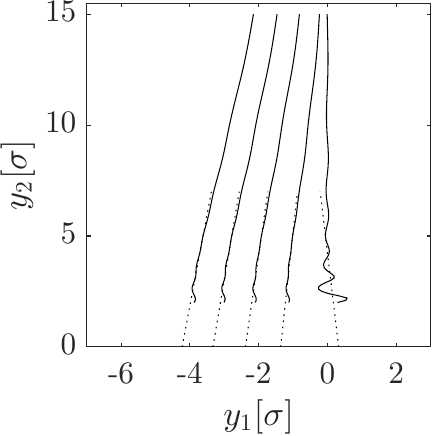}} &
\raisebox{-.5\height}{\includegraphics[height=3cm]{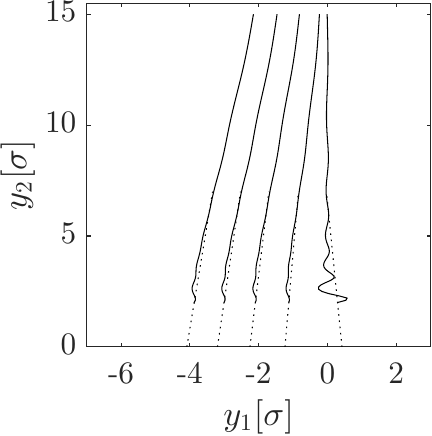}} &
\raisebox{-.5\height}{\includegraphics[height=3cm]{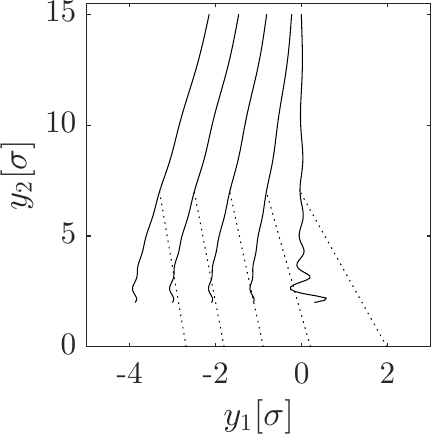}} \\\\
\rotatebox[origin=c]{90}{\textcolor{red}{$h = 6\sigma$}} &
\raisebox{-.5\height}{\includegraphics[height=3cm]{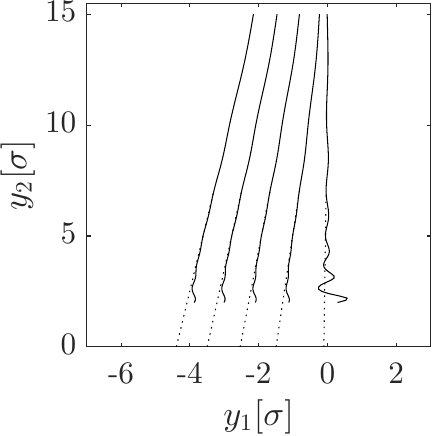}} &
\raisebox{-.5\height}{\includegraphics[height=3cm]{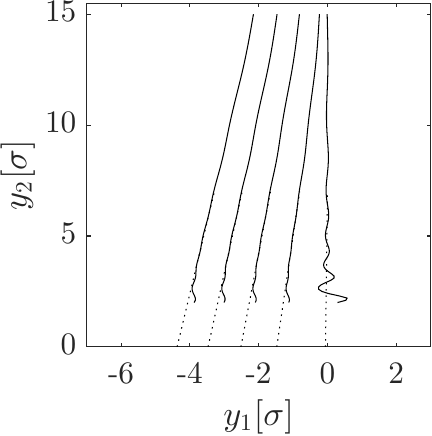}} &
\raisebox{-.5\height}{\includegraphics[height=3cm]{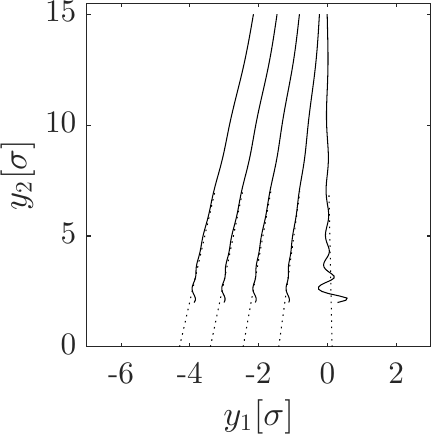}} \\\\
\rotatebox[origin=c]{90}{\textcolor{blue}{$h = 8\sigma$}} &
\raisebox{-.5\height}{\includegraphics[height=3cm]{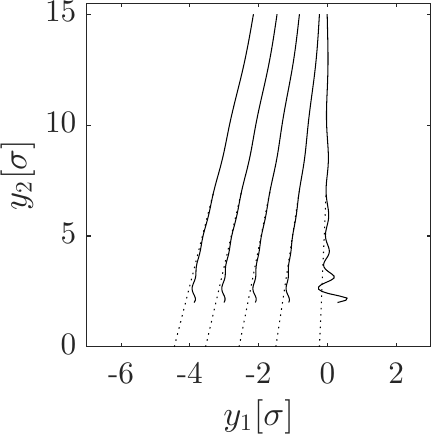}} &
\raisebox{-.5\height}{\includegraphics[height=3cm]{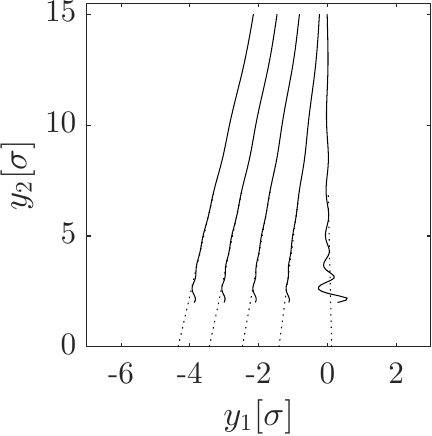}} &
\raisebox{-.5\height}{\includegraphics[height=3cm]{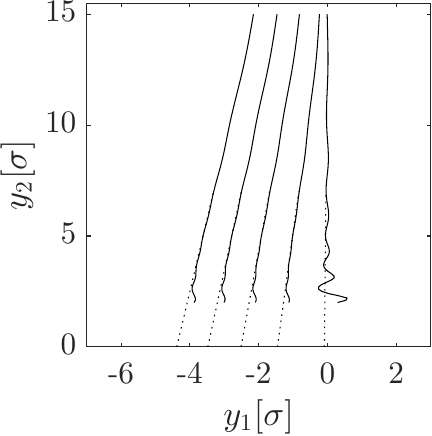}} \\\\
\rotatebox[origin=c]{90}{\textcolor{magenta}{$h = 10\sigma$}} &
\raisebox{-.5\height}{\includegraphics[height=3cm]{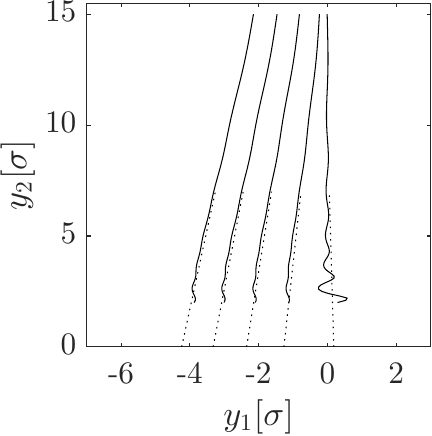}} &
\raisebox{-.5\height}{\includegraphics[height=3cm]{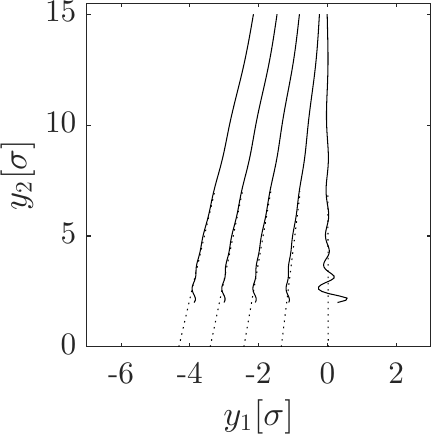}} &
\raisebox{-.5\height}{\includegraphics[height=3cm]{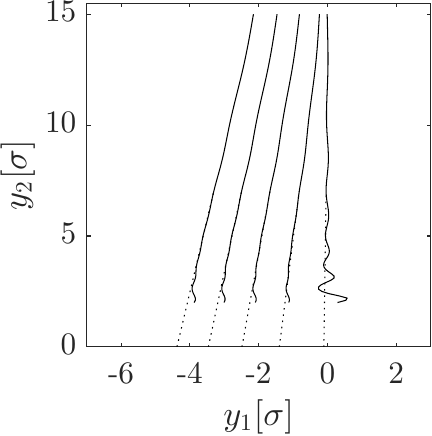}}
\end{tabular}
\begin{tabular}{c}
    \includegraphics[height=3cm]{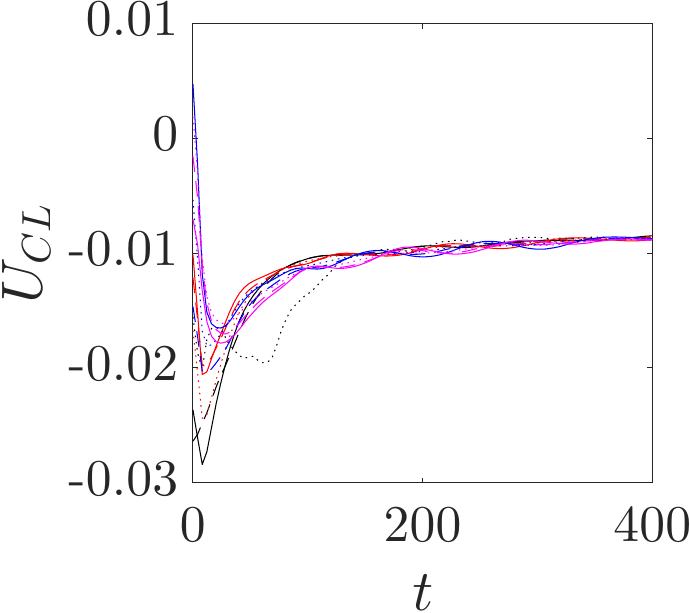} \\
    \includegraphics[height=3cm]{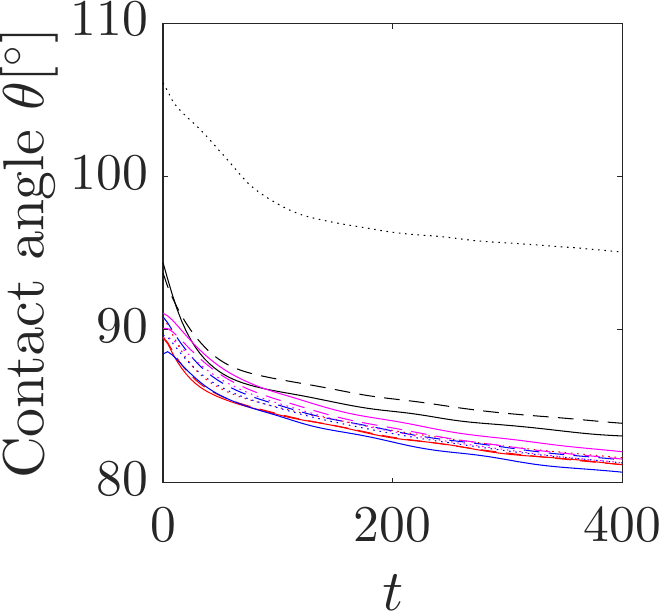} \\
    \includegraphics[height=3cm]{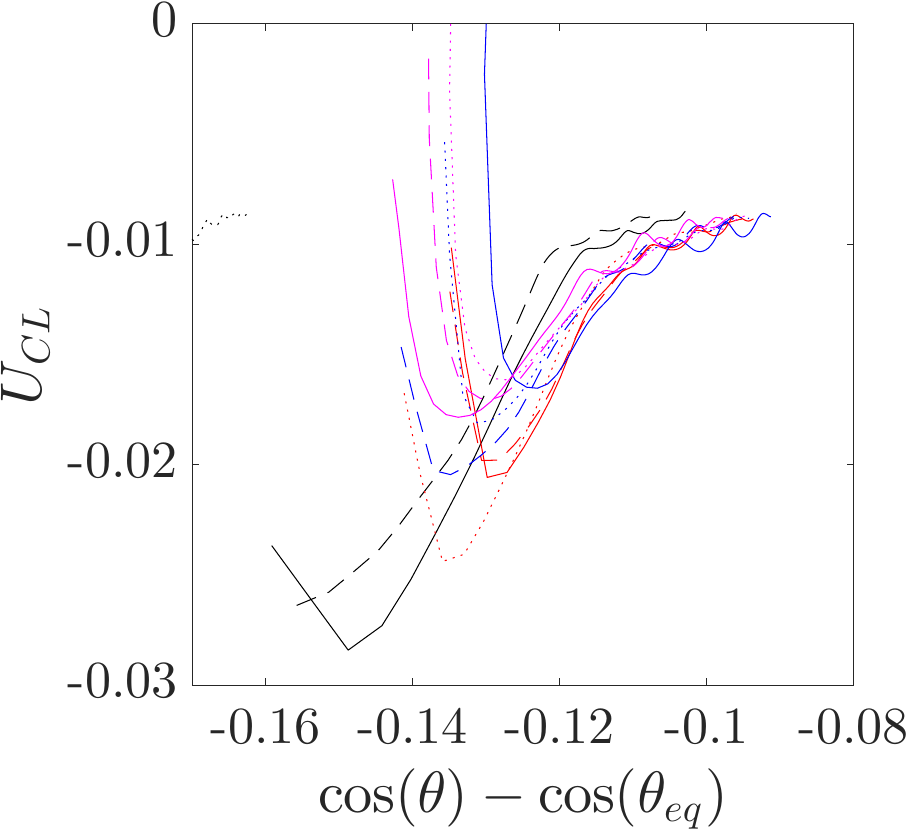} \\
\end{tabular}
\caption{Contact line friction for an advancing contact line, $90^\circ \to 60^\circ$ at $k_B T = 0.75\varepsilon$. Left panels:
    Linear interpolation (dotted lines) of the liquid-vapor interface (black lines). Different lines correspond to the interface at timepoints $t = 0,100,200,300,400\tau$.
    The dotted lines are obtained by interpolating the respective solid line in an interval of width $w$ located at distance $h$ from the wall, $\mathcal{I} = [h-w/2 , h+ w/2]$
    (the size of interval $\mathcal{I}$ was discussed in Appendix~\ref{compdetails}).
    Right panels: Contact line velocity over time, contact angle over time, and crossplot of the two to extract the contact line friction.
    Black, red, blue, and magenta lines encode intervals at distance $h$ equal to $4,6,8$, and $10\sigma$ (corresponding to rows of the left panel).
    Solid, dashed, and dotted lines represent intervals of width $w = 2,4,6\sigma$ (corresponding to columns of the left panel).
    \label{fig:MKT:Robustness:adv:kbT075}}
\end{figure}

\begin{figure}[h!tp]
\centering
\begin{tabular}{c|ccc}
 & $w =2\sigma$ (-) & $w = 4\sigma$ (--\,--) & $w = 6\sigma$ ($\cdot\cdot\cdot$) \\ \hline
\rotatebox[origin=c]{90}{$h = 4\sigma$} &
\raisebox{-.5\height}{\includegraphics[height=3cm]{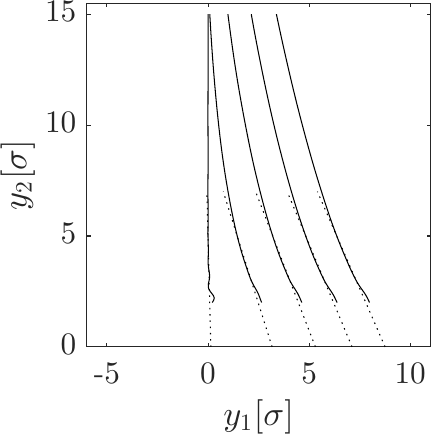}} &
\raisebox{-.5\height}{\includegraphics[height=3cm]{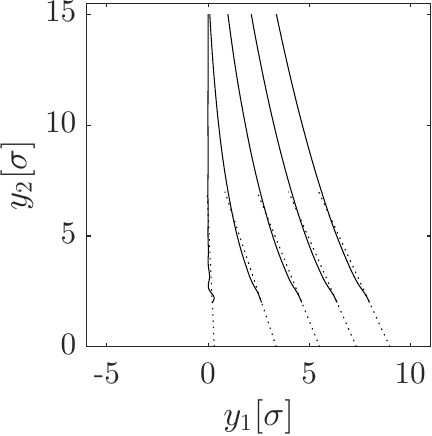}} &
\raisebox{-.5\height}{\includegraphics[height=3cm]{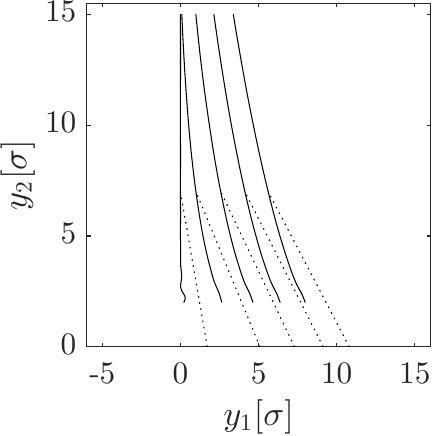}} \\\\
\rotatebox[origin=c]{90}{\textcolor{red}{$h = 6\sigma$}} &
\raisebox{-.5\height}{\includegraphics[height=3cm]{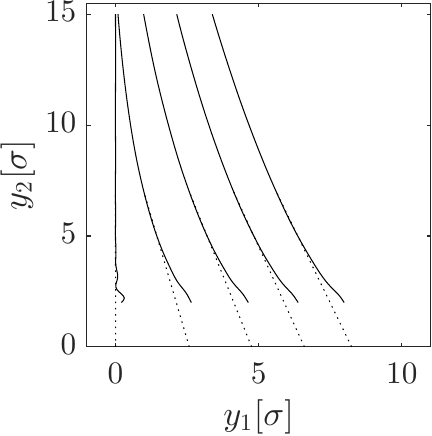}} &
\raisebox{-.5\height}{\includegraphics[height=3cm]{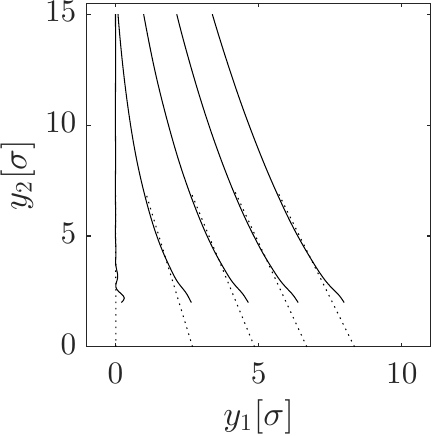}} &
\raisebox{-.5\height}{\includegraphics[height=3cm]{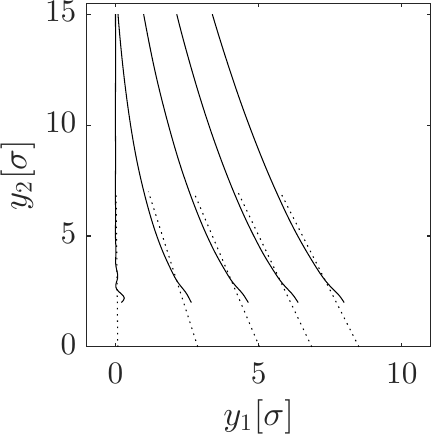}} \\\\
\rotatebox[origin=c]{90}{\textcolor{blue}{$h = 8\sigma$}} &
\raisebox{-.5\height}{\includegraphics[height=3cm]{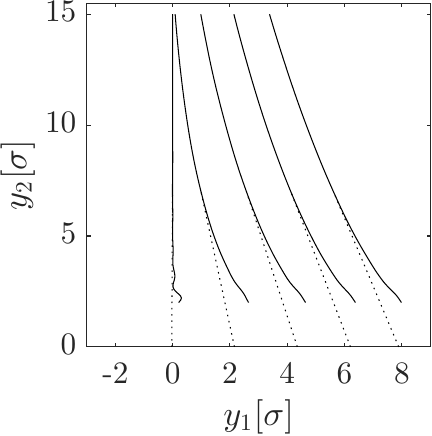}} &
\raisebox{-.5\height}{\includegraphics[height=3cm]{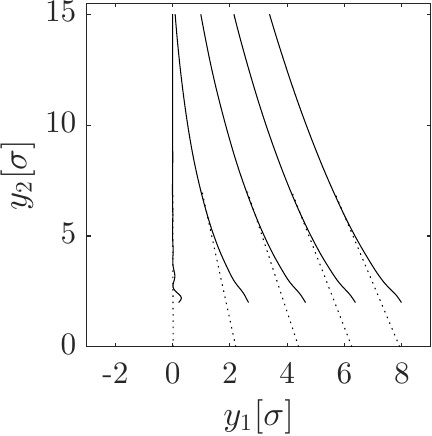}} &
\raisebox{-.5\height}{\includegraphics[height=3cm]{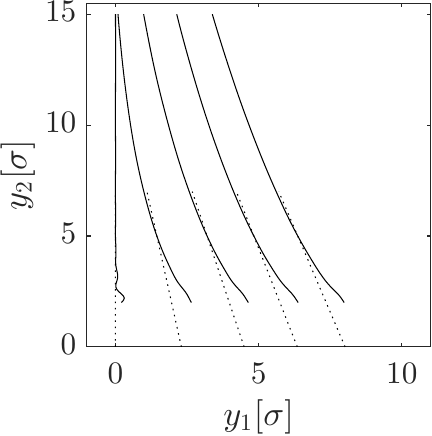}} \\\\
\rotatebox[origin=c]{90}{\textcolor{magenta}{$h = 10\sigma$}} &
\raisebox{-.5\height}{\includegraphics[height=3cm]{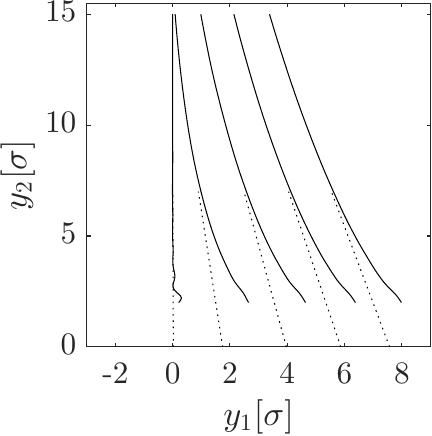}} &
\raisebox{-.5\height}{\includegraphics[height=3cm]{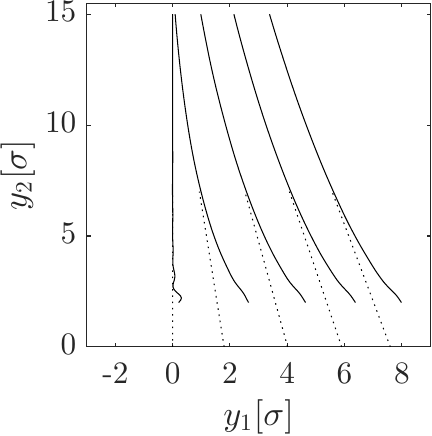}} &
\raisebox{-.5\height}{\includegraphics[height=3cm]{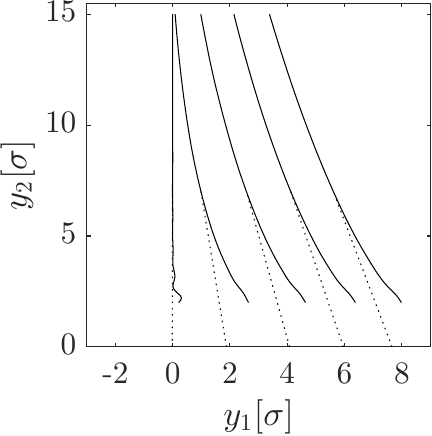}}
\end{tabular}
\begin{tabular}{c}
    \includegraphics[height=3cm]{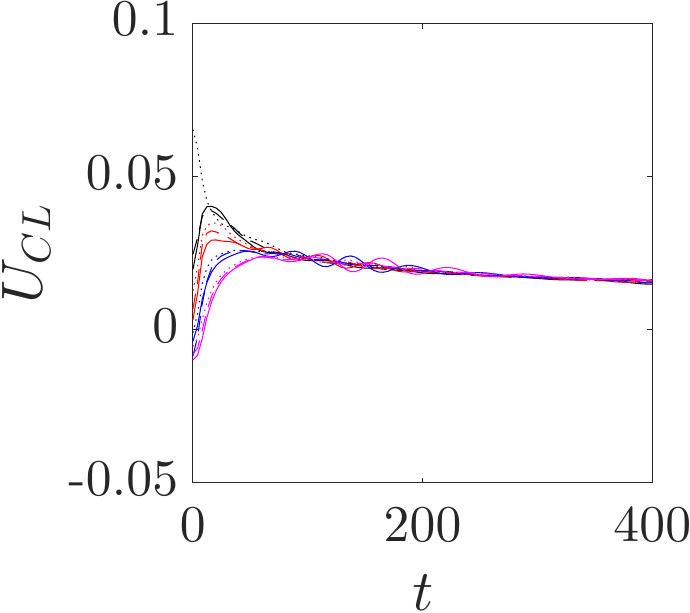} \\
    \includegraphics[height=3cm]{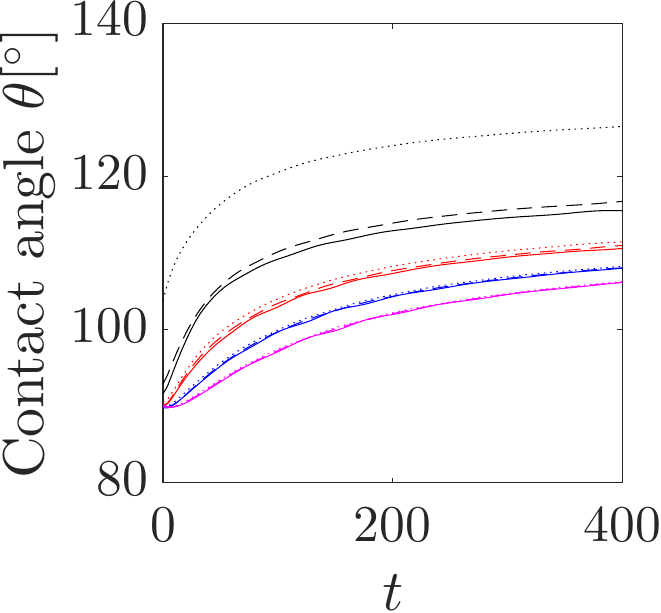} \\
    \includegraphics[height=3cm]{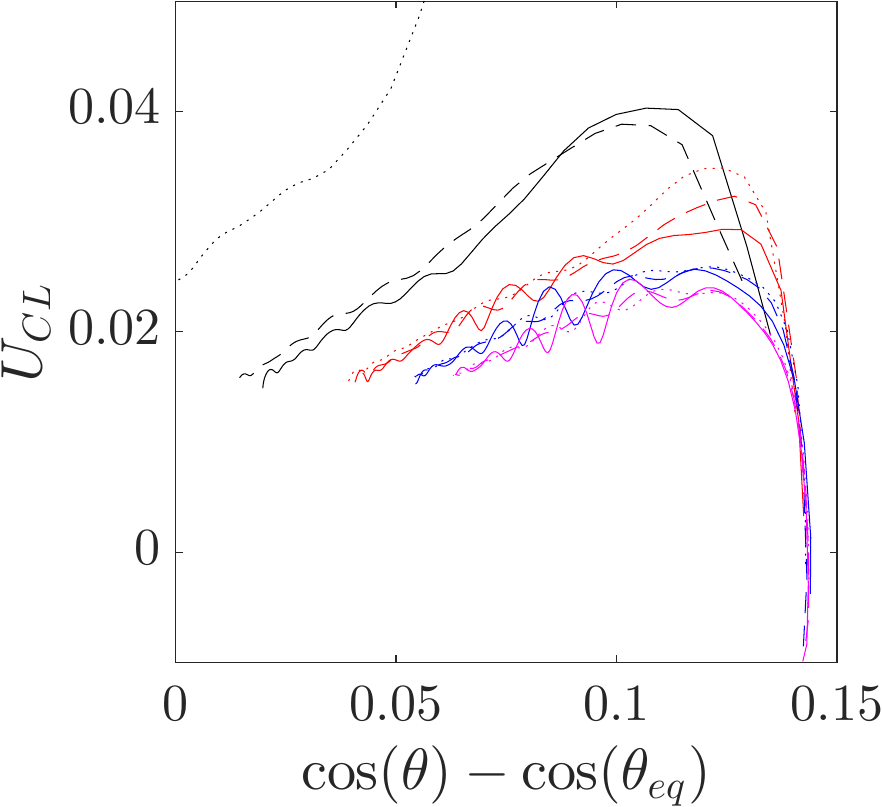} \\
\end{tabular}
\caption{Contact line friction for a receding contact line, $90^\circ \to 120^\circ$ at $k_B T = 0.75\varepsilon$. Details as in Fig.~\ref{fig:MKT:Robustness:adv:kbT075}.
\label{fig:MKT:Robustness:red:kbT075}}
\end{figure}

\begin{figure}[h!tp]
\centering
\begin{tabular}{c|ccc}
 & $w =2\sigma$ (-) & $w = 4\sigma$ (--\,--) & $w = 6\sigma$ ($\cdot\cdot\cdot$) \\ \hline
\rotatebox[origin=c]{90}{$h = 4\sigma$} &
\raisebox{-.5\height}{\includegraphics[height=3cm]{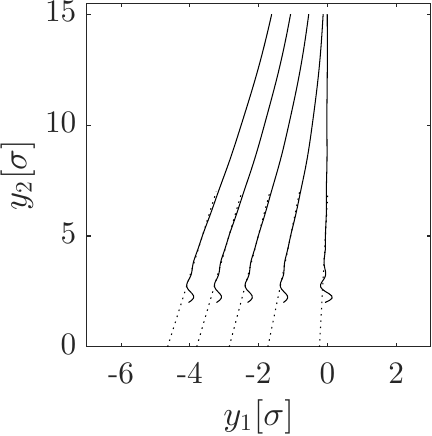}} &
\raisebox{-.5\height}{\includegraphics[height=3cm]{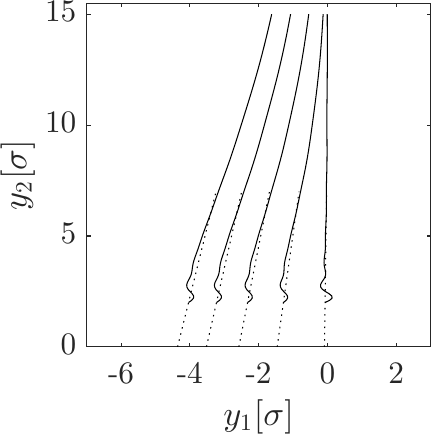}} &
\raisebox{-.5\height}{\includegraphics[height=3cm]{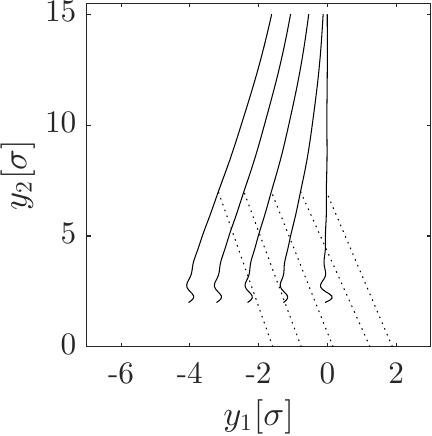}} \\\\
\rotatebox[origin=c]{90}{\textcolor{red}{$h = 6\sigma$}} &
\raisebox{-.5\height}{\includegraphics[height=3cm]{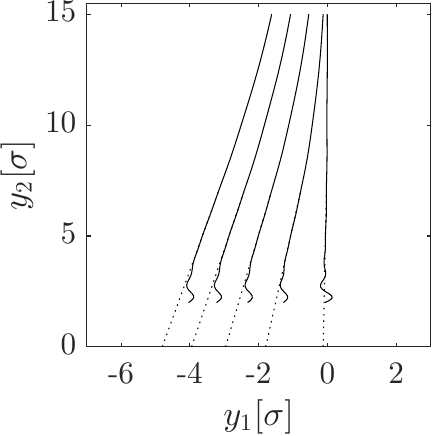}} &
\raisebox{-.5\height}{\includegraphics[height=3cm]{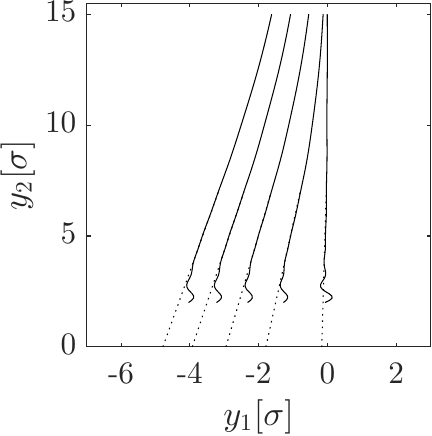}} &
\raisebox{-.5\height}{\includegraphics[height=3cm]{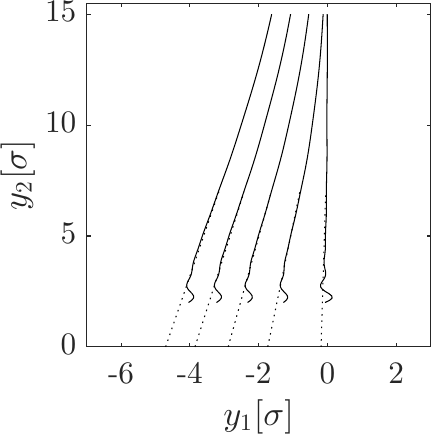}} \\\\
\rotatebox[origin=c]{90}{\textcolor{blue}{$h = 8\sigma$}} &
\raisebox{-.5\height}{\includegraphics[height=3cm]{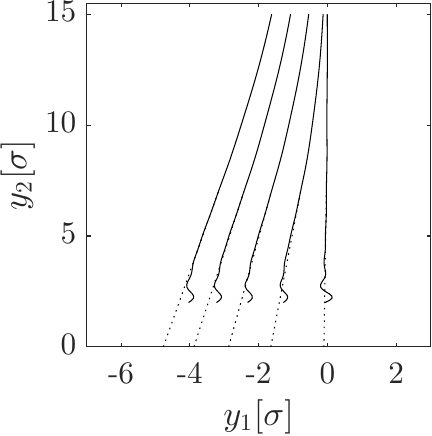}} &
\raisebox{-.5\height}{\includegraphics[height=3cm]{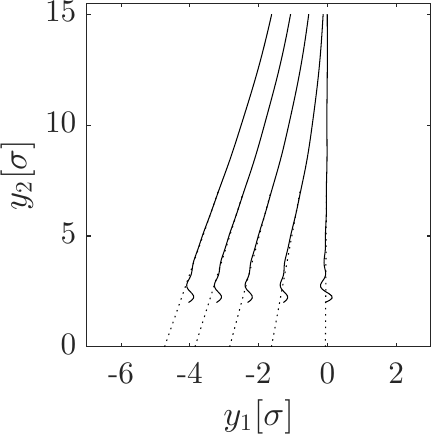}} &
\raisebox{-.5\height}{\includegraphics[height=3cm]{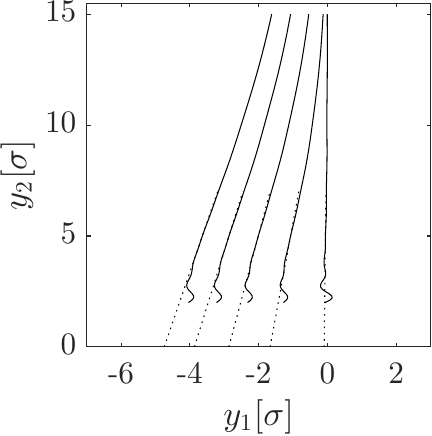}} \\\\
\rotatebox[origin=c]{90}{\textcolor{magenta}{$h = 10\sigma$}} &
\raisebox{-.5\height}{\includegraphics[height=3cm]{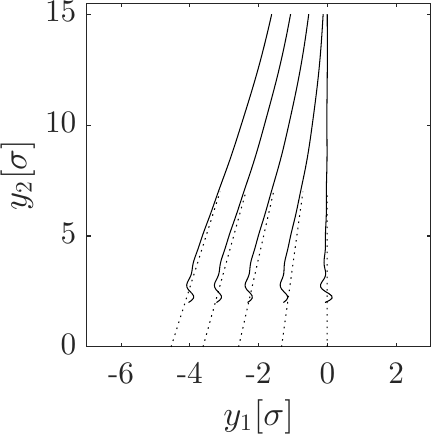}} &
\raisebox{-.5\height}{\includegraphics[height=3cm]{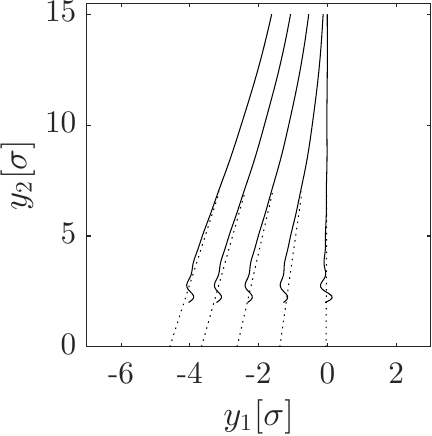}} &
\raisebox{-.5\height}{\includegraphics[height=3cm]{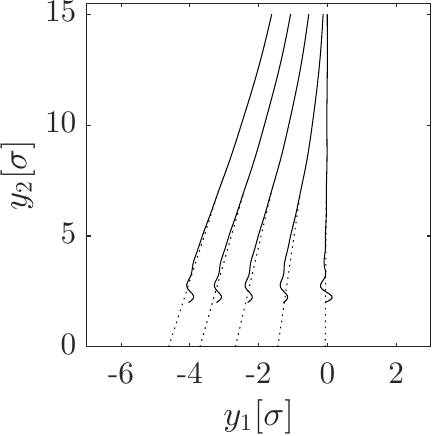}}
\end{tabular}
\begin{tabular}{c}
    \includegraphics[height=3cm]{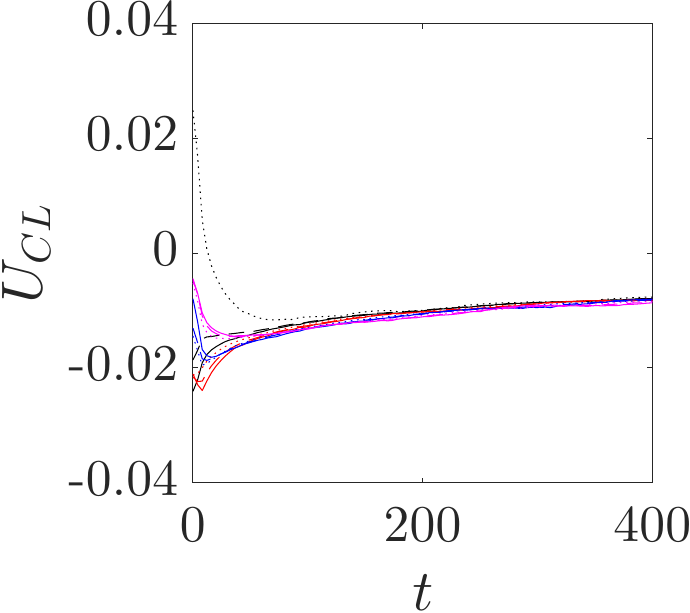} \\
    \includegraphics[height=3cm]{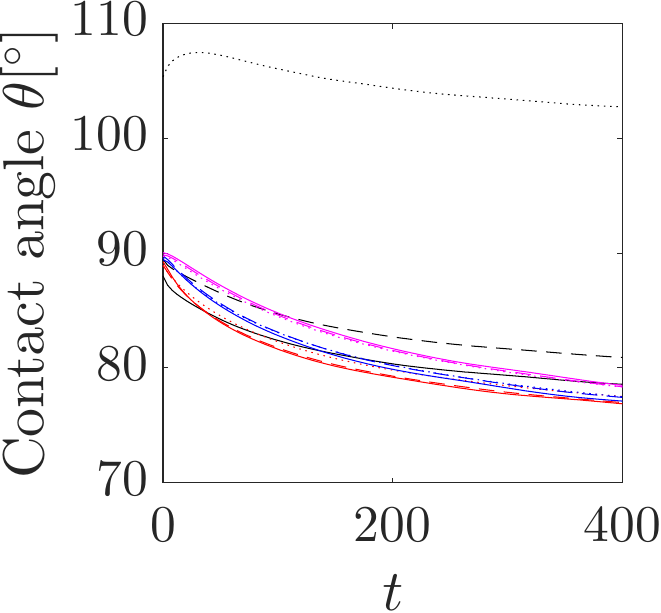} \\
    \includegraphics[height=3cm]{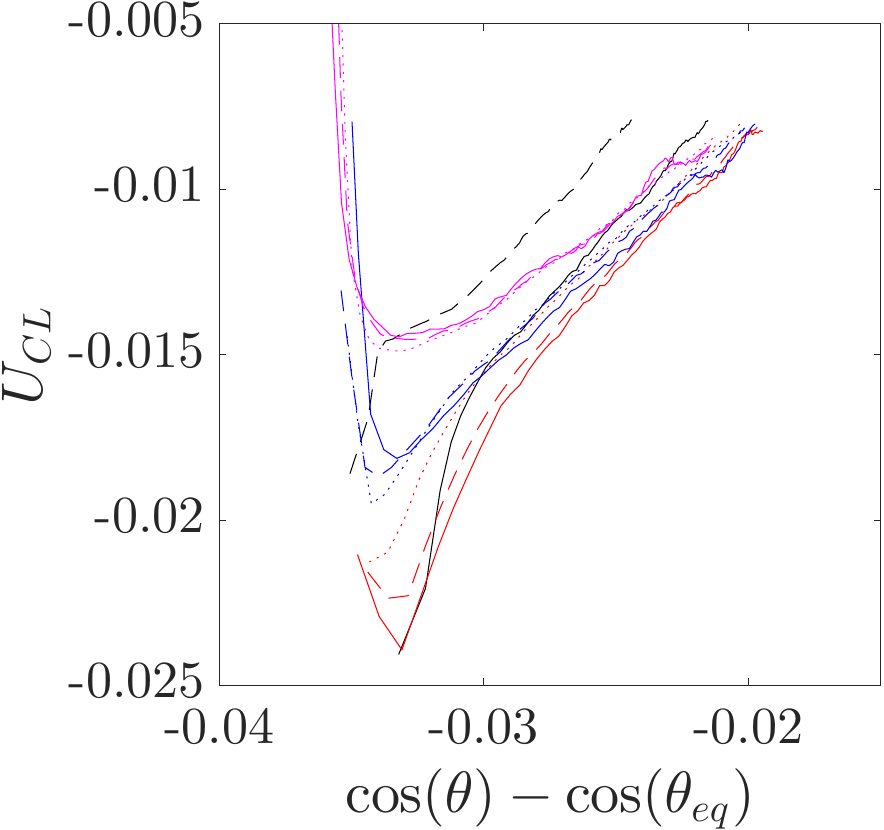} \\
\end{tabular}
\caption{Contact line friction for an advancing contact line, $90^\circ \to 60^\circ$ at $k_B T = 0.9\varepsilon$. Details as in Fig.~\ref{fig:MKT:Robustness:adv:kbT075}.
\label{fig:MKT:Robustness:adv:kbT09}}
\end{figure}

\begin{figure}[h!tp]
\centering
\begin{tabular}{c|ccc}
 & $w =2\sigma$ (-) & $w = 4\sigma$ (--\,--) & $w = 6\sigma$ ($\cdot\cdot\cdot$) \\ \hline
\rotatebox[origin=c]{90}{$h = 4\sigma$} &
\raisebox{-.5\height}{\includegraphics[height=3cm]{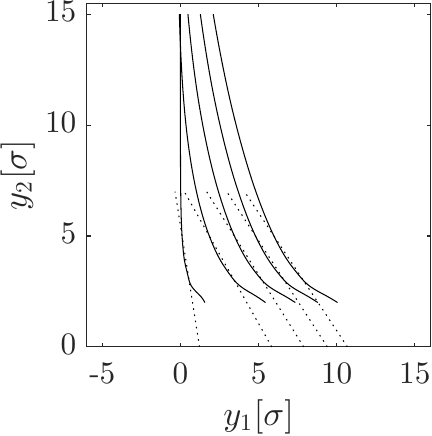}} &
\raisebox{-.5\height}{\includegraphics[height=3cm]{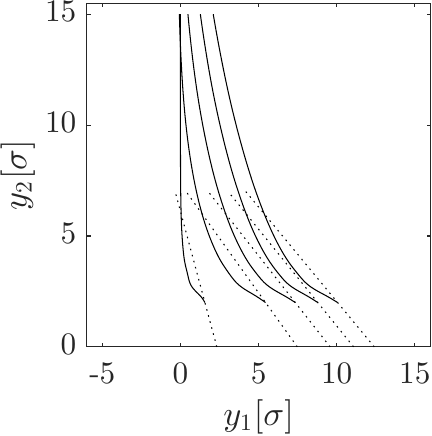}} &
\raisebox{-.5\height}{\includegraphics[height=3cm]{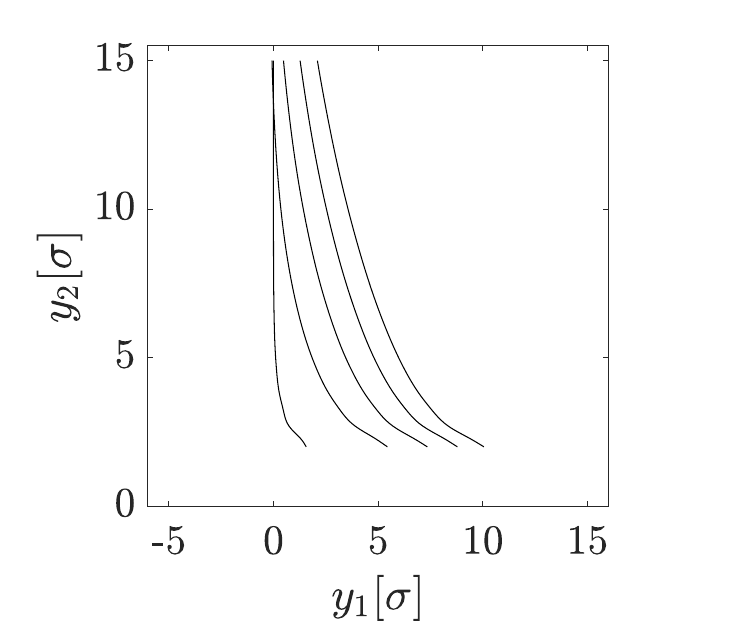}} \\\\
\rotatebox[origin=c]{90}{\textcolor{red}{$h = 6\sigma$}} &
\raisebox{-.5\height}{\includegraphics[height=3cm]{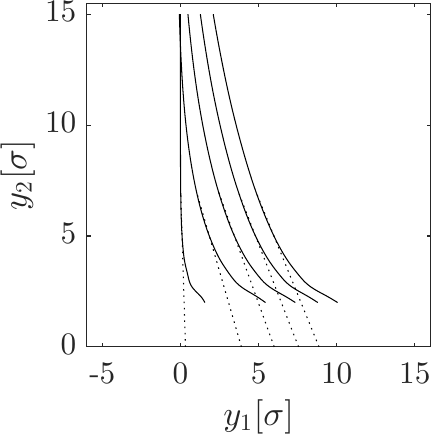}} &
\raisebox{-.5\height}{\includegraphics[height=3cm]{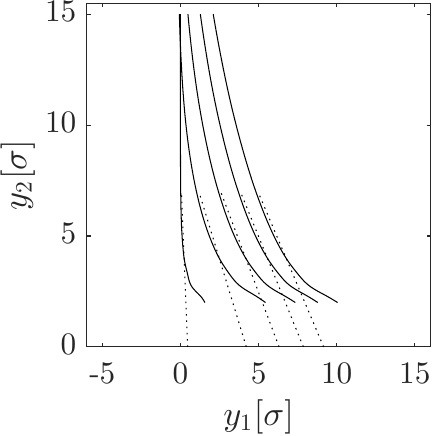}} &
\raisebox{-.5\height}{\includegraphics[height=3cm]{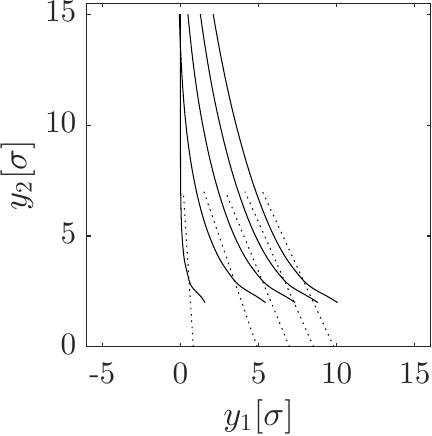}} \\\\
\rotatebox[origin=c]{90}{\textcolor{blue}{$h = 8\sigma$}} &
\raisebox{-.5\height}{\includegraphics[height=3cm]{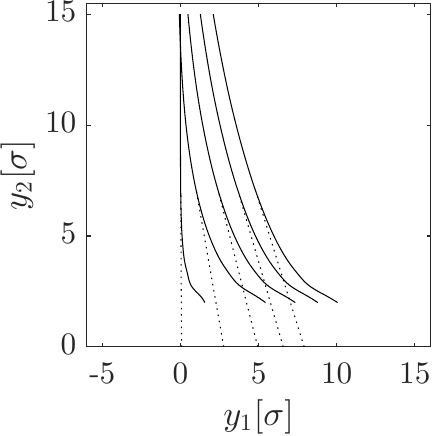}} &
\raisebox{-.5\height}{\includegraphics[height=3cm]{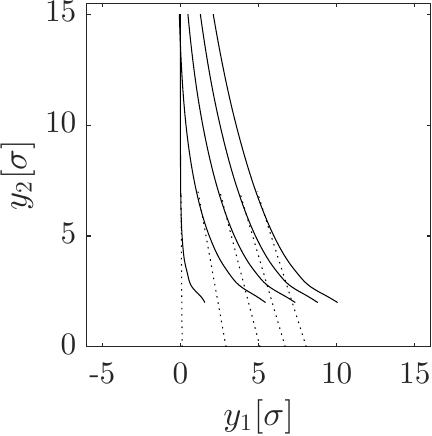}} &
\raisebox{-.5\height}{\includegraphics[height=3cm]{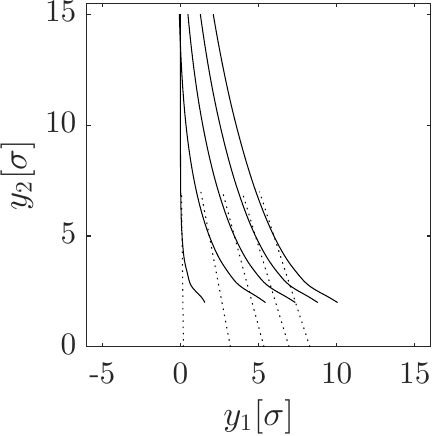}} \\\\
\rotatebox[origin=c]{90}{\textcolor{magenta}{$h = 10\sigma$}} &
\raisebox{-.5\height}{\includegraphics[height=3cm]{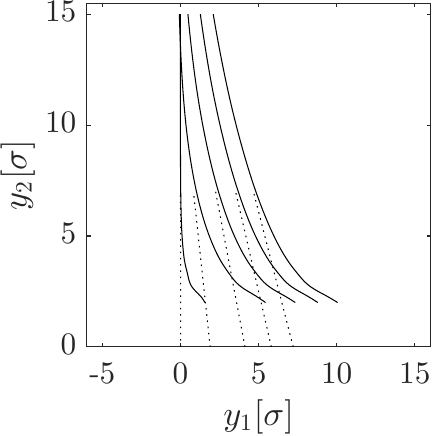}} &
\raisebox{-.5\height}{\includegraphics[height=3cm]{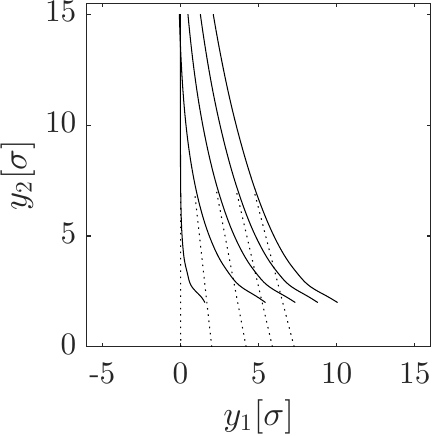}} &
\raisebox{-.5\height}{\includegraphics[height=3cm]{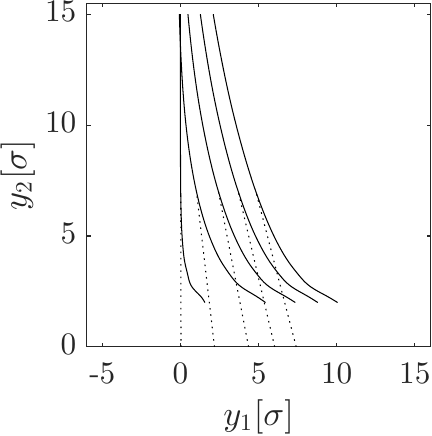}}
\end{tabular}
\begin{tabular}{c}
    \includegraphics[height=3cm]{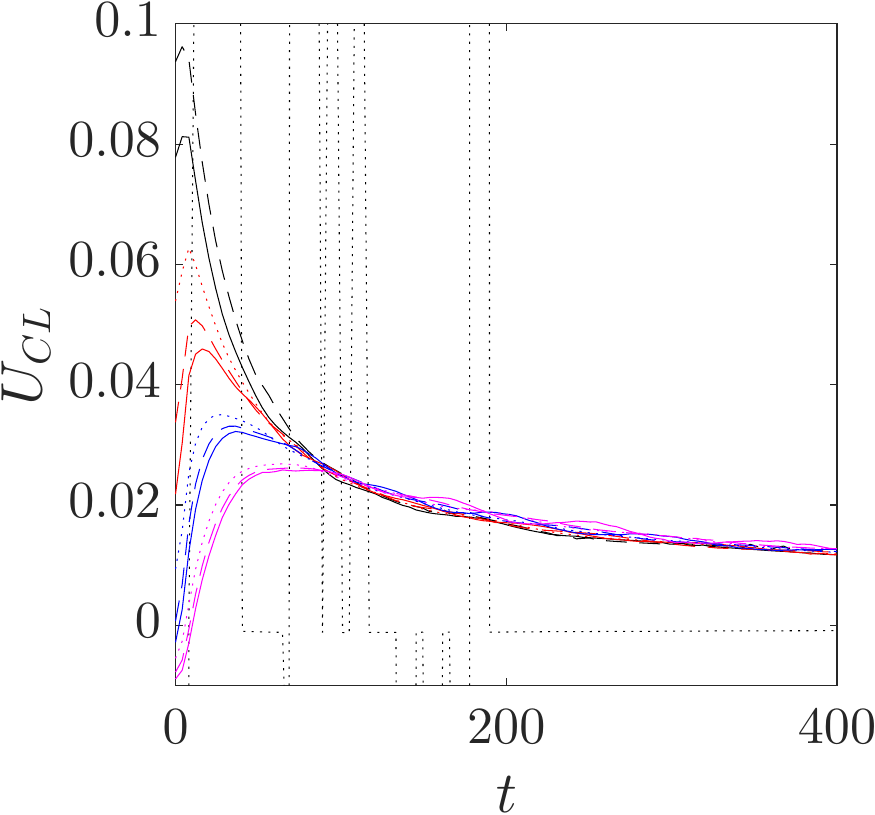} \\
    \includegraphics[height=3cm]{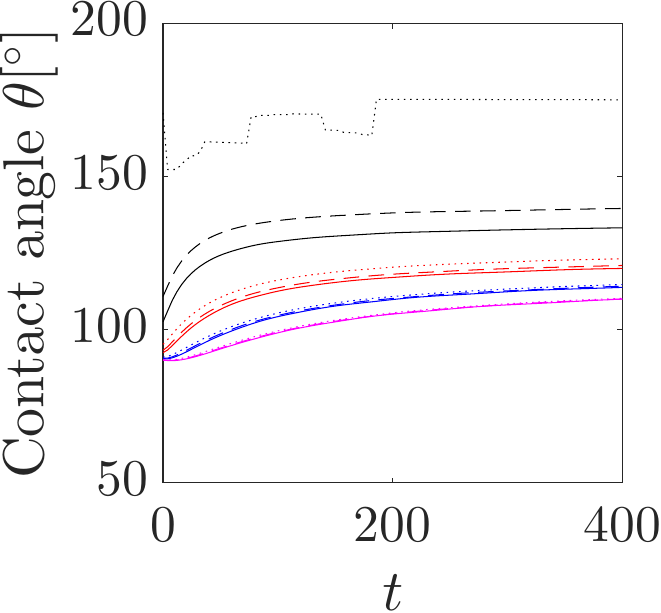} \\
    \includegraphics[height=3cm]{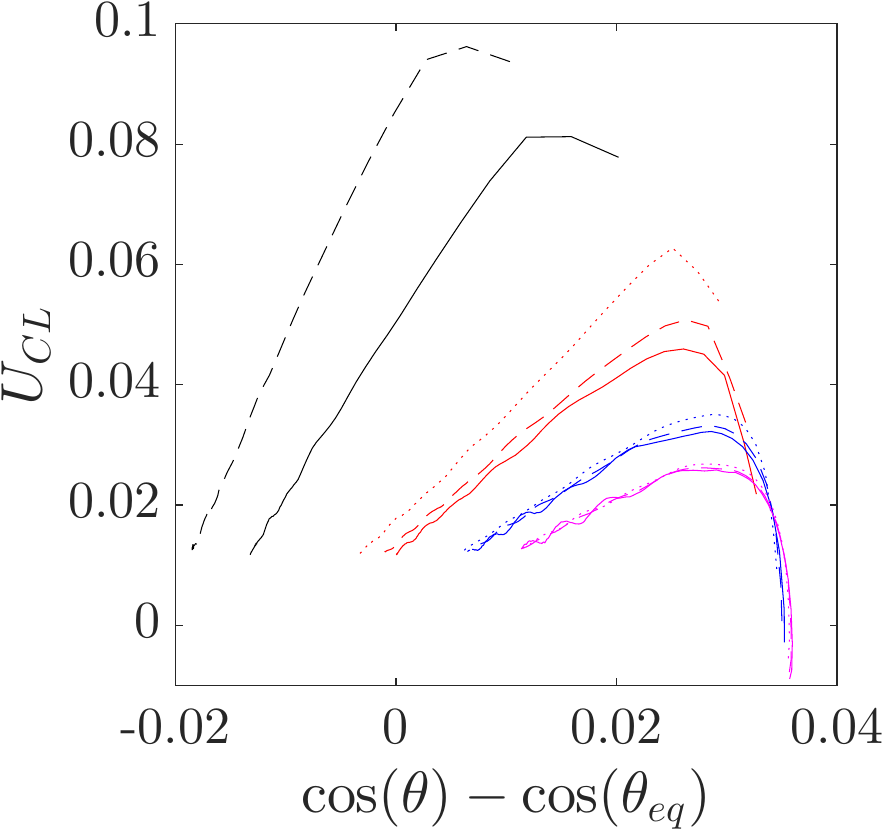} \\
\end{tabular}
\caption{Contact line friction for a receding contact line, $90^\circ \to 120^\circ$ at $k_B T = 0.9\varepsilon$. Details as in Fig.~\ref{fig:MKT:Robustness:adv:kbT075}. Note that for $h=4\sigma$ and $w = 6\sigma$, the interface extends along the wall in the interval $\mathcal{I} = [1\sigma ,7\sigma]$, therefore not allowing for an effective fit.
\label{fig:MKT:Robustness:rec:kbT09}}
\end{figure}

\begin{figure}[h!tp]
\centering
\includegraphics{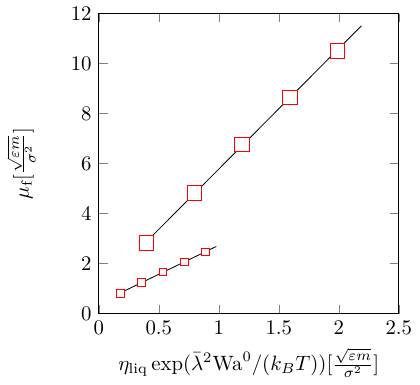}
\includegraphics{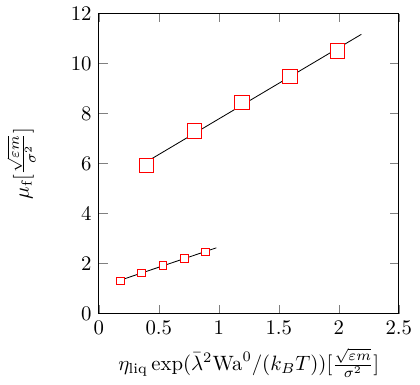}
\caption{Contact line friction measurements for varying viscosities, for an advancing contact line with $\thYoung = 70^\circ$ and initial contact angle
$\thInitial = 90^\circ$ for $k_BT = 0.75 \varepsilon$ and  $0.9\varepsilon$ represented by large and small symbols, respectively.
Left: All viscosities, i.e. bulk and shear viscosity for liquid and vapor phases, were scaled by $[0.2,0.4,0.6,0.8,1.0]$.
Right: Only the liquid shear viscosity is scaled by $[0.2,0.4,0.6,0.8,1.0]$. Black lines are linear fits to the data. Scaling the liquid shear viscosity by $0.2$
leads to a residual contact line friction of approximately $50\%$ of its original value. \label{fig:ChangeOfContactLineFrictionWithViscosity}}
\end{figure}

\begin{figure}[h!tp]
\centering
\includegraphics{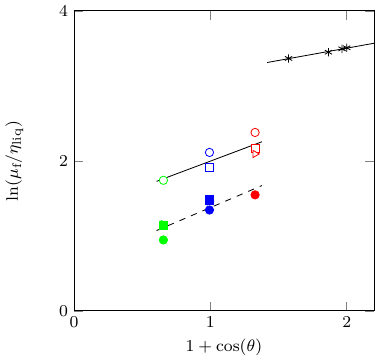}
\caption{Comparison of the contact line friction as in Fig.~\ref{fig:5} for $k_BT = 0.75\varepsilon$, with MD simulations Seveno and De Coninck~\cite{seveno2004possibility} of
capillary rise around a fiber of 16-atom-long liquid molecules, represented by $\ast$-symbols. Black lines are linear fits to the data.
\label{fig:contactLineFriction_ComparisonWithExperiment}}
\end{figure}

Figures~\ref{fig:MKT:Robustness:adv:kbT075}-\ref{fig:MKT:Robustness:rec:kbT09}
present contact line friction computations illustrating the robustness of
this parameter. Figure~\ref{fig:ChangeOfContactLineFrictionWithViscosity}
illustrates the effect of shear viscosity on contact line friction and
Fig.~\ref{fig:contactLineFriction_ComparisonWithExperiment} reports a
comparison with MD.

\newcommand{\AdvLowTemp}{2019_1_23_9_54_7}
\newcommand{\RecLowTemp}{2019_1_23_10_20_27}
\newcommand{\AdvHighTemp}{2019_1_16_12_22_34}
\newcommand{\RecHighTemp}{2019_1_16_12_37_58}

\newpage
\bibliography{Bibliography}

\end{document}